\documentclass[nofootinbib,aps,prd,reprint,superscriptaddress]{revtex4-2}

\usepackage[utf8]{inputenc} 
\usepackage{graphicx,overpic,mathtools}
\usepackage{amsthm,amsmath,amssymb,hyperref}
\usepackage{braket,bm,bbm,setspace}
\usepackage[normalem]{ulem} 
\usepackage{physics}
\usepackage{float}
\usepackage[makeroom]{cancel}
\usepackage[english]{babel}
\usepackage{xcolor}
\usepackage{tensor}
\graphicspath{ {images/} }
\addto\captionsspanish{}
\hypersetup{
	colorlinks=true,
	pdfborder={0 0 0},
	citecolor=purple,
	linkcolor=blue,
	filecolor=blue,
	urlcolor=blue,
}
\usepackage{subfigure}

\usepackage{orcidlink}

\usepackage{booktabs}
\usepackage{multirow}

\usepackage{pdfcomment}

\makeatletter
\def\l@subsection#1#2{}
\def\l@subsubsection#1#2{}
\makeatother

\begin{document}
\title{Randomized SearchRank:\\ A Semiclassical Approach to a Quantum Search Engine}
\author{Sergio A. Ortega}
\email{sergioan@ucm.es}
\affiliation{Departamento de Física Teórica, Universidad Complutense de Madrid, 28040 Madrid, Spain}
\author{Miguel A. Martin-Delgado}
\email{mardel@ucm.es}
\affiliation{Departamento de Física Teórica, Universidad Complutense de Madrid, 28040 Madrid, Spain}
\affiliation{CCS-Center for Computational Simulation, Universidad Politécnica de Madrid, 28660 Boadilla del Monte, Madrid, Spain.}

\begin{abstract}
{The quantum SearchRank algorithm is a promising tool for a future quantum search engine based on PageRank quantization. However, this algorithm loses its functionality when the $N/M$ ratio between the network size $N$ and the number of marked nodes $M$ is sufficiently large. We propose a modification of the algorithm, replacing the underlying Szegedy quantum walk with a semiclassical walk. To maintain the same time complexity as the quantum SearchRank algorithm we propose a simplification of the algorithm. This new algorithm is called Randomized SearchRank, since it corresponds to a quantum walk over a randomized mixed state. The performance of the SearchRank algorithms is first analyzed on an example network, and then statistically on a set of different networks of increasing size and different number of marked nodes. On the one hand, to test the search ability of the algorithms, it is computed how the probability of measuring the marked nodes decreases with $N/M$ for the quantum SearchRank, but remarkably it remains at a high value around $0.9$ for our semiclassical algorithms, solving the quantum SearchRank problem. The time complexity of the algorithms is also analyzed, obtaining a quadratic speedup with respect to the classical ones. On the other hand, the ranking functionality of the algorithms has been investigated, obtaining a good agreement with the classical PageRank distribution. Finally, the dependence of these algorithms on the intrinsic PageRank damping parameter has been clarified. Our results suggest that this parameter should be below a threshold so that the execution time does not increase drastically.}
\end{abstract}

\maketitle

\section{Introduction}\label{Introduction}

The PageRank algorithm was a revolution in the field of search engines for surfing the Internet \cite{Brin1,Brin2,Brin3,Google_book}. This algorithm is able to rank pages objectively, taking into account the structure of the network formed by them. Moreover, beyond its importance in the World Wide Web, it has a multitude of applications, such as in bibliometrics \cite{PR-Biblio-1,PR-Biblio-2}, finance \cite{PR-Finances}, metabolic networks \cite{PR-Metabolism}, drug discovery \cite{PR-Drug}, protein interaction networks \cite{PR-BQ} and social networks \cite{PR-Twitter}. With the advent of quantum networks for a future quantum Internet, a quantum version of PageRank was devised in 2012 \cite{Paparo1}. This algorithm was originally intended to rank information from quantum networks, as it was expected to give more sensible results than the classical algorithm. Moreover, it has also proven to be a valuable technique for classical information, outperforming the classical PageRank results \cite{Paparo2,APR}. Since its development, there has been a great deal of research interest in this algorithm, including experimental realizations \cite{DTQW-PR,CT-QPR,Comparing-CQ,TF_QPR,Exp_QPR}.

The quantum PageRank algorithm relies on the Szegedy's quantum walk \cite{Szegedy}. The classical simulation of this quantum walk belongs to the computational complexity class $P$ \cite{Squwals}. Thus, the quantum PageRank can be simulated efficiently in a classical computer, without the need of a fault-tolerant quantum computer. Despite this, classifying the information is not the only task that a search engine performs. It also has to search for the pages of interest, providing them to the user. Since there exist quantum algorithms for searching problems that outperforms classical ones, as for example the Grover algorithm \cite{Grover}, in 2014 a new quantum algorithm that integrates the quantum search into the quantum PageRank was devised. It was dubbed quantum SearchRank, and it was the first quantum algorithm able to search the nodes of interest at the same time that provides a ranking for them \cite{Searchrank}. Moreover, it was the first algorithm implementing a Szegedy's quantum walk with queries to an oracle, with a quadratic speedup. Later, different formulations of Szegedy's quantum walk with queries were analyzed in the field of quantum search \cite{S_queries}.

\begin{figure*}
	\centering
	\includegraphics[scale=0.2375]{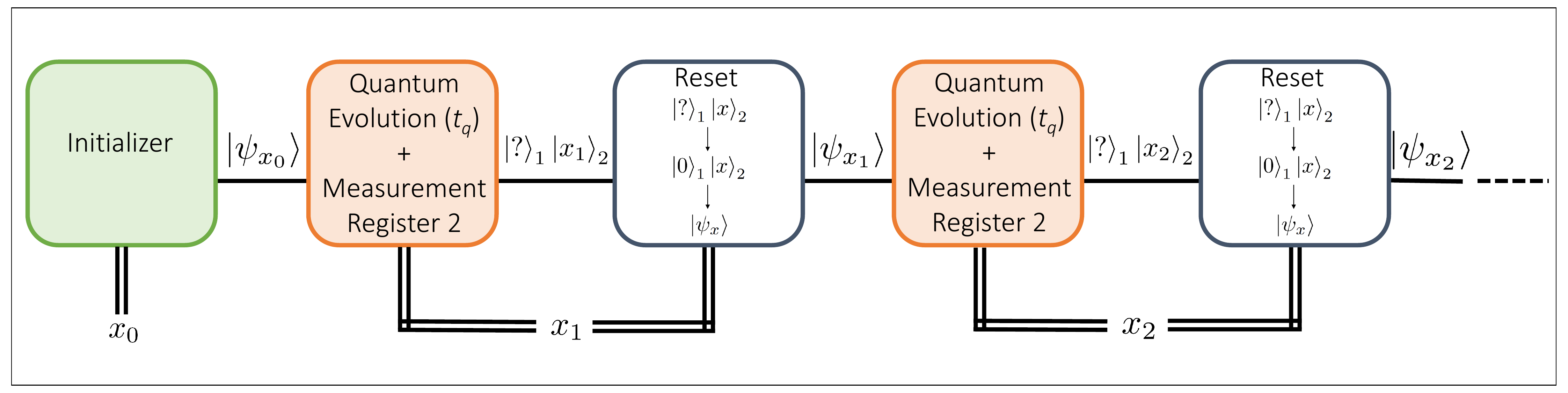}
	\caption{Semiclassical Szegedy's walk of class II. Let us denote $x_{t_c}$ as the position of the walker at classical time step $t_c$. Thus, we start at node $x_0$. We prepare the proxy state $\left|\psi_{x_0}\right>$ in \eqref{psi_i} and perform the quantum evolution, parameterized by the quantum time $t_q$ being the number of applications of the quantum walk evolution operator. After measuring the second register the system collapses to a particular node $x_1$ in the second register. The remaining information in the first register, represented by a question mark, plays no role in the algorithm, so we do not worry about it. Before proceeding to the next step, the system must be reset. To this end, the first register is erased, so that it is forced non-unitarily to be in $\left|0\right>_1$. After that, we use the measured information about the node $x_1$ to prepare with a suitable unitary evolution the new proxy state, $\left|\psi_{x_1}\right>$ in \eqref{psi_i}, completing the reset of the system. This constitutes a classical step of the semiclassical walk, and the process is repeated the number of classical steps $t_c$ as desired.}
	\label{F:class_II}
\end{figure*}

The quantum SearchRank algorithm seems a promising tool for a future quantum search engine. However, it has some problems that need to be solved. One of them is that the search functionality seems to break down when the size of the network is large enough, so that the nodes of interest are not found correctly. In this work we propose a semiclassical approach for the quantum SearchRank, based on the recently developed semiclassical Szegedy's walk \cite{Semiclassical}. In addition, we provide a simplification to speed up the semiclassical walk, giving rise to a quantum algorithm that we refer to as Randomized SearchRank. This algorithm performs a quantum search from a mixed-state, which has been previously studied in the context of Grover's algorithm \cite{Grover_mixed_1,Grover_mixed_2}. As we will show throughout this paper, we are able to measure the nodes of interest with a high probability regardless of the size of the problem, at the same time that the time complexity of the quantum SearchRank is preserved. Another missing issue in the quantum SearchRank algorithm is a lack of statistical analysis about the performance of the ranking functionality, which we will tackle properly in this work. We will show that the SearchRank algorithms provide a ranking compatible with the classical PageRank, and thus our Randomized SearchRank is useful for sampling this distribution with a quadratic speedup. Finally, we shall also analyze how the damping parameter of the PageRank algorithms affects to the SearchRank, obtaining a maximum threshold for our semiclassical approach to be useful.

This paper is organized as follows. In Section \ref{Background} we review the formulation of the PageRank and the quantum SearchRank algorithms, as well as the semiclassical Szegedy's walk. In Section \ref{Sc-SearchRank} we introduce the Semiclassical SearchRank algorithm. In Section \ref{Example} we apply the SearchRank algorithms to an example of scale-free graph. In Section \ref{Searching} we focus on the searching feature of the SearchRank algorithms, analyzing the time complexity and the amplification of the probability. In Section \ref{Ranking} we focus on the ranking feature of the SearchRank algorithms. In Section \ref{Alpha} we study the dependence with the damping parameter inherent to the PageRank algorithm. Finally, we summarize and conclude in Section \ref{Conclusions}.

\section{Background}\label{Background}

For completeness and further reference in this work, in this section we will briefly describe the PageRank \cite{Paparo1} and SearchRank \cite{Searchrank} algorithms, as well as the semiclassical walks \cite{Semiclassical} recently introduced. Further details can be found in the original references.

\begin{figure*}
	\centering
	\includegraphics[width=1.0\linewidth]{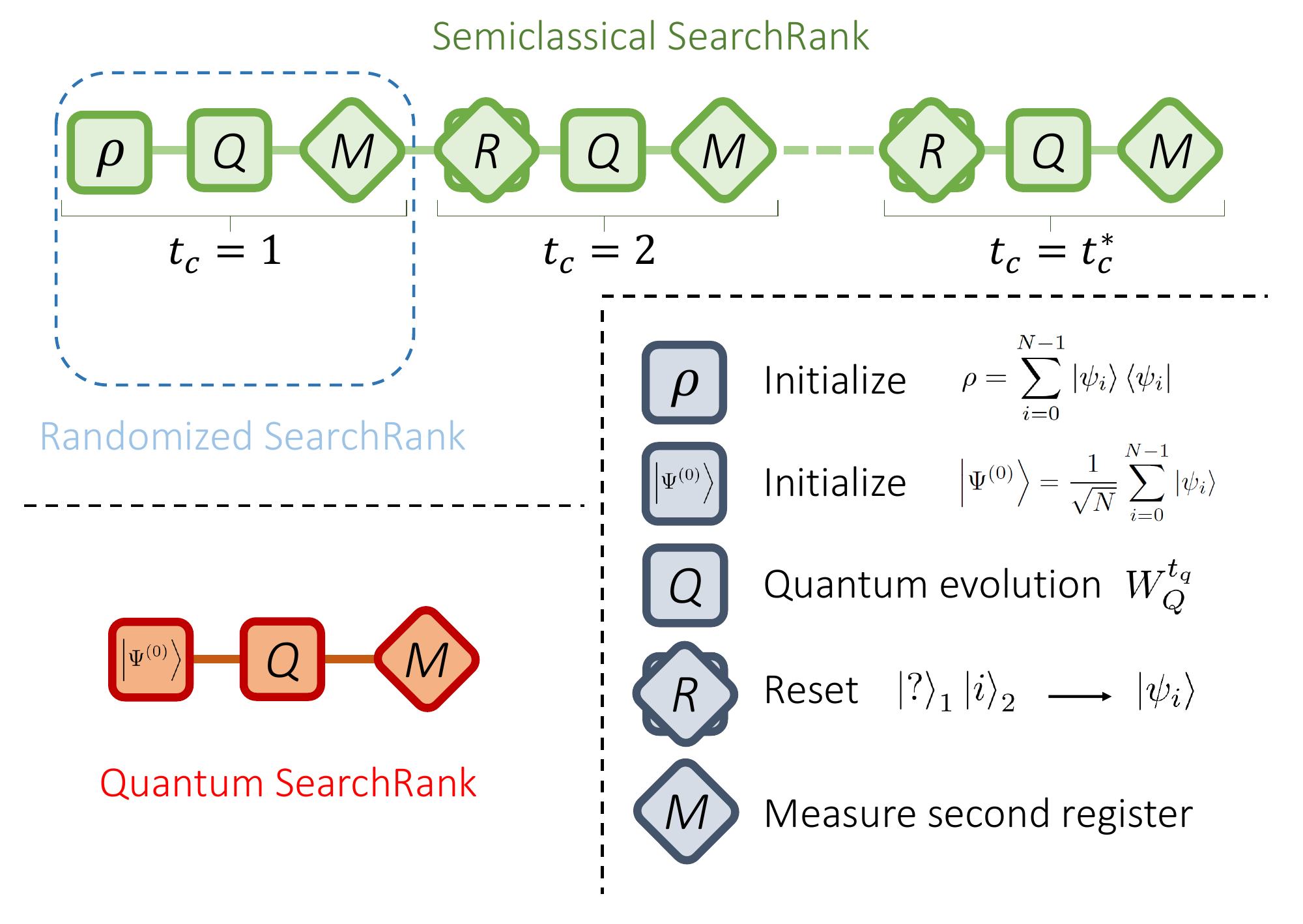}
	\caption{Quantum circuit diagrams of SearchRank algorithms. In the Semiclassical SearchRank (upper panel in green), the first step consists on the initialization of the mixed state $\rho$, the quantum evolution of $t_q$ times the unitary operator $W_Q$, and a measurement in the second register. After that, each classical step consists on a reset of the system depending on the previous measurement, the quantum evolution, and a measurement. In total, $t_c^*$ classical steps are carried out until convergence. The blue dashed box represents the Randomized SearchRank, which is a simplified semiclassical algorithm with only one classical step. In the quantum SearchRank (bottom left panel in red), the initial state $\left|\Psi^{(0)}\right>$ is prepared, the quantum evolution is performed and the system is measured. The right dashed box is a legend explaining the meaning of the different elements (quantum circuits) from which the SearchRank algorithms are constructed. In particular, notice that the reset operation is a combination of a measurement and evolution operation.}
	\label{F:semiclassical}
\end{figure*}

\subsection{Classical and Quantum PageRank}

The PageRank algorithm provides a method to classify the pages of the internet by importance. These pages form a network, where we can define the associated connectivity $N\times N$ matrix $H$ as:
\begin{equation}
	H_{i,j} := 
	\left\lbrace\begin{array}{c}
		1/\text{outdeg}(P_j) \ \ \ \text{if} \ P_j \in B_i,\\
		0 \ \ \ \ \ \ \ \ \ \ \ \ \ \ \ \ \ \ \ \text{otherwise},
	\end{array}
	\right.
\end{equation} 
where $B_i$ is the set of nodes linking to the node $P_i$, outdeg($P_j$) is the outdegree of the node $P_j$, and $N$ is the number of nodes in the network.

In order to compute the PageRank, the matrix $H$ is patched so that each null column is filled with $1/N$, giving rise to matrix $E$. This is a column-stochastic matrix, where all columns sum up to one. Next, this matrix $E$ is mixed with another matrix \textbf{1} where all entries are equal to $1$, obtaining a primitive and irreducible matrix called the Google matrix $G$:
\begin{equation}\label{G}
	G := \alpha E + \frac{(1-\alpha)}{N} \text{\textbf{1}}.
\end{equation}
The parameter $\alpha$ corresponding to the previous mixing is called the damping parameter, and its value lies in $[0,1]$. It was found by Brin and Page that the optimal value is $\alpha = 0.85$. This is the value that is used to compute both the classical and quantum PageRank.

The classical PageRank vector $I_c$ is defined as the stationary distribution of the Google matrix, i.e., $GI_c = I_c$. This vector is obtained with a power method, applying repeatedly the matrix $G$ to an initial probability distribution. Given the way $G$ is constructed, this method ends up converging to the PageRank distribution.

The power method of the classical algorithm is equivalent to a random walk where the transition matrix is the Google matrix $G$. Thus, the quantum PageRank algorithm is based on Szegedy's quantum walk, which is a general quantization of a Markov chain. The Hilbert space is the span of all the vectors representing the $N \times N$ directed edges of the duplicated graph, i.e., $\mathcal{H} = \text{span}\lbrace\left|i\right>_1\left|j\right>_2,\ i,j \in N \times N\rbrace = \mathbb{C}^N \otimes \mathbb{C}^N$, where the states with indexes $1$ and $2$ refers to the nodes on two copies of the original graph. In this paper we count the nodes of the network, and therefore the matrix indexes, from $0$ to $N-1$. We define the vectors:
\begin{equation}\label{psi_i}
	\left|\psi_i\right> := \left|i\right>_1 \otimes \sum_{k=0}^{N-1} \sqrt{G_{ki}}\left|k\right>_2,
\end{equation}
which are a superposition of the vectors representing the edges outgoing
from the $i^{th}$ vertex, whose coefficients are given by the square root of the $i^{th}$ column of the matrix $G$. From these vectors we define the projector operator onto the subspace generated by them:
\begin{equation}
	\Pi := \sum_{k=0}^{N-1} \left|\psi_k\right>\left<\psi_k\right|.
\end{equation}
The quantum walk operator $U$ is defined as:
\begin{equation}\label{U}
	U := SR,
\end{equation}
where $R$ is a reflection over the subspace generated by the $\left|\psi_i\right>$ states,
\begin{equation}\label{reflection}
	R := 2\Pi - \mathbbm{1},
\end{equation}
and $S$ is the swap operator between the two quantum registers, i.e.,
\begin{equation}
	S := \sum_{i,j=0}^{N-1} \left|i,j\right>\left<j,i\right|.
\end{equation}

The operator $U$ must be applied an even number of times, since the exchange operator changes the network directions, so the real time evolution operator is chosen as $W := U^2$.

The initial state of the quantum system is chosen as the equal superposition of all the $\left|\psi_i\right>$ states,
\begin{equation}\label{initial}
	\left|\Psi^{(0)}\right> := \frac{1}{\sqrt{N}} \sum_{i=0}^{N-1} \left|\psi_i\right>,
\end{equation}
and the information of the quantum PageRank is obtained measuring the second register, since this one contains the information of where the directed links point to. Moreover, due to the unitary character of the evolution, the quantum walk does not converge to a fixed point. For that reason, the quantum PageRank is defined as the average of the instantaneous quantum distributions at different times:
\begin{equation}
	I_q(P_i) := \frac{1}{T} \sum_{t=0}^T \left|\left|\tensor[_2]{\left<i\right|W^t\left|\Psi^{(0)}\right>}{}\right|\right|^2.
\end{equation}
This quantity ends up converging for a large enough value of $T$ \cite{Paparo2}.

\subsection{Quantum SearchRank}\label{quantum_searchrank}

At the core of quantum search algorithms there is the oracle operator $Q$ \cite{Grover,Portugal}. Given a set $\mathcal{M}$ of $M$ nodes to search, this operator marks the corresponding computational basis states inverting their sign. Thus, the action of the oracle is defined as follows:
\begin{equation}
	Q\left|i\right> := 
	\left\lbrace\begin{array}{c}
		-\left|i\right> \ \ \ \ \ \text{if} \ i \in \mathcal{M},\\
		\ \ \ \ \left|i\right> \ \ \ \ \    \text{otherwise}.
	\end{array}
	\right.
\end{equation}

In the case of the quantum SearchRank algorithm, an oracle that marks the nodes in the first register of the Hilbert space is introduced. This will amplify the probability of measuring only the nodes that we are looking for. Let us define this oracle as
\begin{equation}
	Q_1 := Q \otimes \mathbbm{1}_N,
\end{equation}
and modify the unitary evolution operator $U$ introducing the oracle operator between the swap $S$ and the reflection $R$, obtaining $U_Q$:
\begin{equation}
	U_Q := SQ_1R.
\end{equation}
Furthermore, for the quantum SearchRank algorithm the Google matrix used for the $\left|\psi_i\right>$ states in \eqref{psi_i} is constructed with $\alpha = 0.25$ without further ado \cite{Searchrank}. In section \ref{Alpha} we will examine further the values of this parameter and its stability behavior. Finally, as in the quantum PageRank algorithm, unitary  evolution must be applied an even number of times, so the actual unitary evolution operator is $W_Q := U_Q^2$.

The initial state of the system is constructed as before, according to \eqref{initial}, and the second register is measured again to obtain the instantaneous SearchRank distributions at each time step:
\begin{equation}
	S_q(P_i,t) := \left|\left|\tensor[_2]{\left<i\right|W_Q^t\left|\Psi^{(0)}\right>}{}\right|\right|^2.
\end{equation}
Whereas in the quantum PageRank algorithm the different time distributions are averaged to obtain the ranking of the nodes, in this case one is only interested in the distribution where the probability of measuring the marked nodes is amplified. In \cite{Searchrank} it was shown that the optimal time for measuring is approximately $t \approx \sqrt{N/M}$. Later on we will examine this scaling more carefully.

\begin{figure}[htpb]
	\centering
	\includegraphics[scale=0.5]{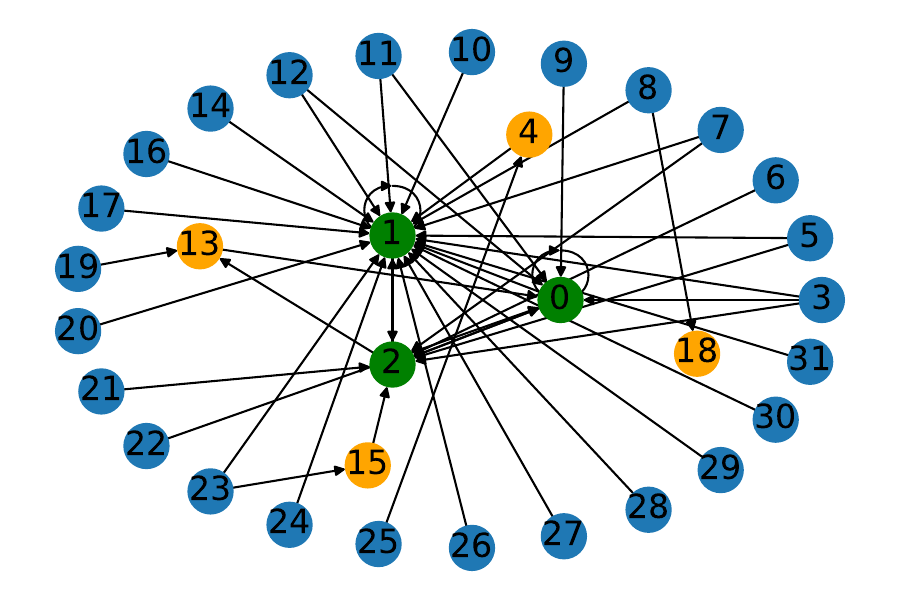}
	\caption{Scale-free network with 32 nodes. The inner (green) nodes correspond to the main nodes. The middle (orange) nodes correspond to secondary nodes. The outer (blue) nodes correspond to residual nodes without links pointing to them.}
	\label{F:graph}
\end{figure}

\begin{figure*}[htpb]
	\centering
	\includegraphics[scale=0.7]{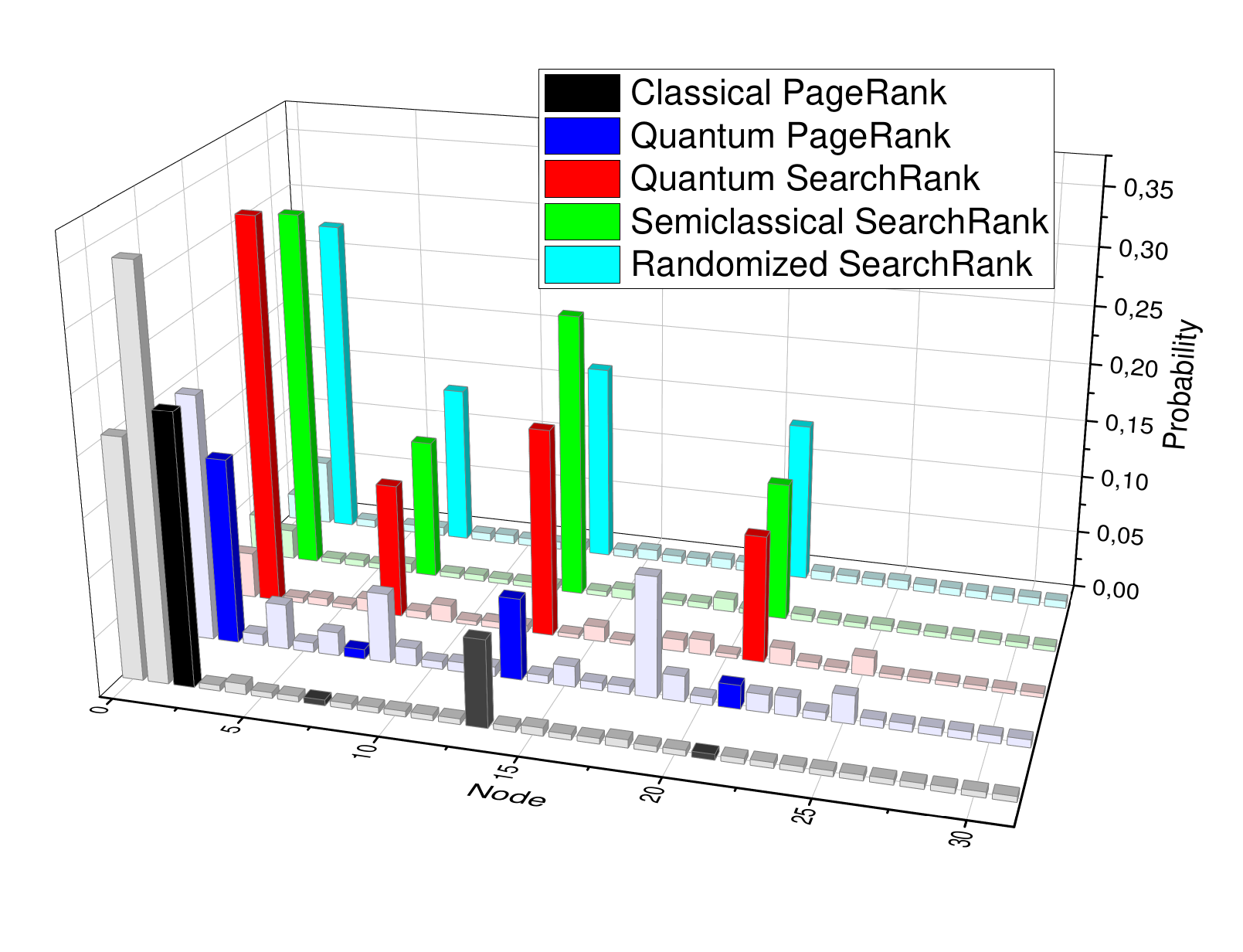}
	\caption{PageRank and SearchRank distributions of the network with 32 nodes of Figure \ref{F:graph}. The marked nodes (2, 7, 13 and 21) have a highlighted color. In the three SearchRank algorithms the marked nodes have an amplified importance.}
	\label{F:histogram}
\end{figure*}

\subsection{Szegedy's Semiclassical Walk}

Semiclassical walks are a new paradigm that combines both classical and quantum features. From a functional point of view in a quantum computer, these walks consist on repeated measurements of the walker position at regular intervals of time. There are two parameters to describe a semiclassical walk:
\newpage
\begin{itemize}
\item Quantum time \cite{QT} $t_q$, which is the number of times we apply a unitary evolution $U$ between measurements.
\item Classical time $t_c$, which is the number of times that the scheme of quantum evolution and measurement is repeated.
\end{itemize}

\begin{figure}[htpb]
	\centering
	\includegraphics[scale=0.5]{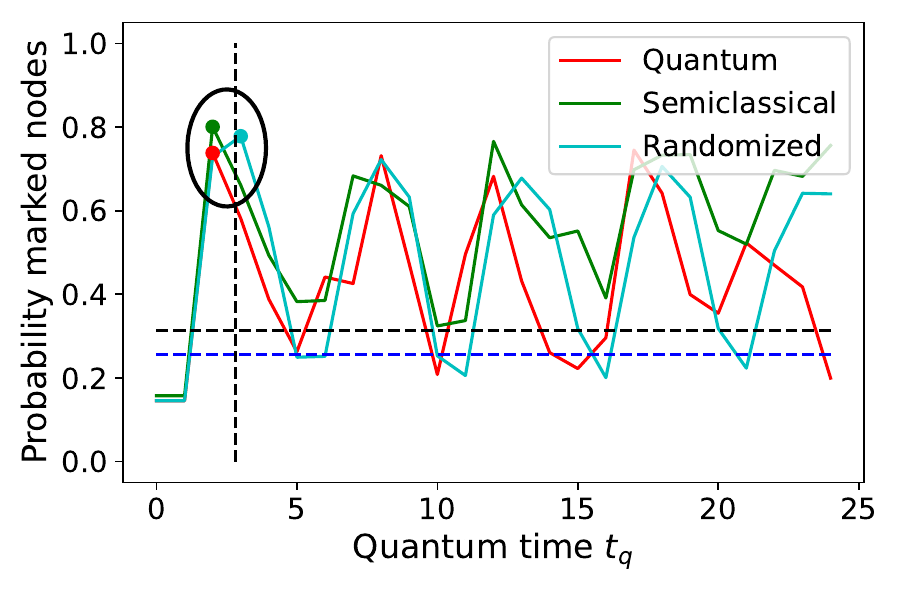}
	\caption{Probability of measuring one of the marked nodes versus the quantum time for the three SearchRank algorithms applied in the scale-free graph with 32 nodes of Figure \ref{F:graph}. The first maximum of each curve is marked with a dot, and they are surrounded by a circle. The vertical dashed line represents the reference time $\sqrt{N/M}$. The horizontal dashed lines represent the probability of the marked nodes in the classical (black) and quantum (blue) PageRank distributions.}
	\label{F:probabilities}
\end{figure}

\begin{figure*}
	\makebox[0pt][c]{
		\subfigure[]{\includegraphics[scale=0.45]{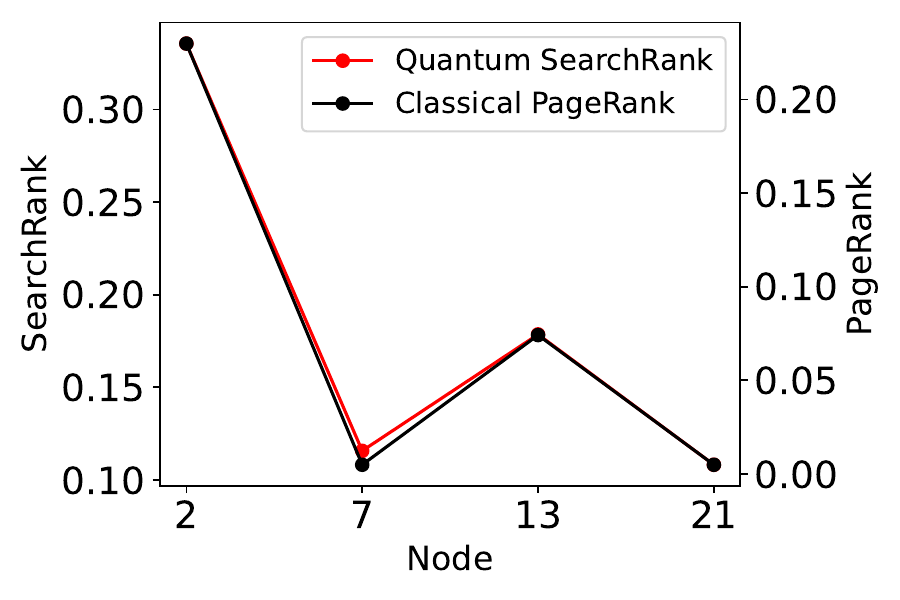}\label{F:classification_C_Q}}
		\hspace{-6pt}
		\subfigure[]{\includegraphics[scale=0.45]{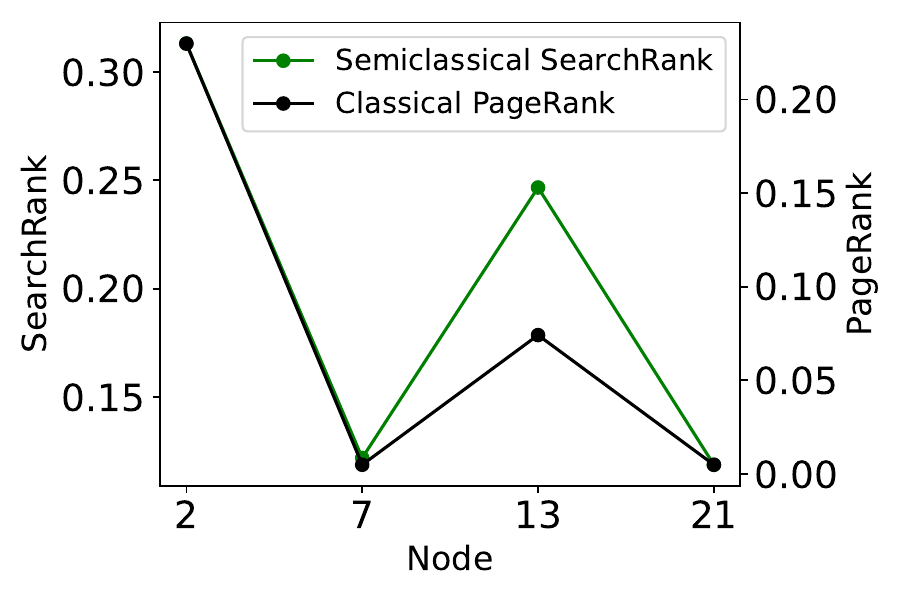}\label{F:classification_C_S}}
		\hspace{-6pt}
		\subfigure[]{\includegraphics[scale=0.45]{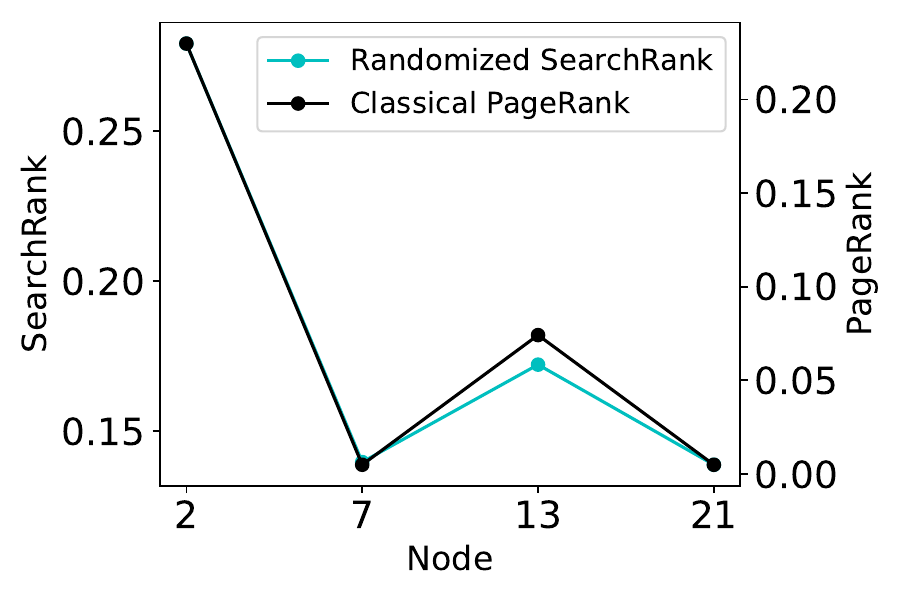}\label{F:classification_C_Si}}
	}
	\\
	\vspace{-7pt}
	\makebox[0pt][c]{
		\subfigure[]{\includegraphics[scale=0.45]{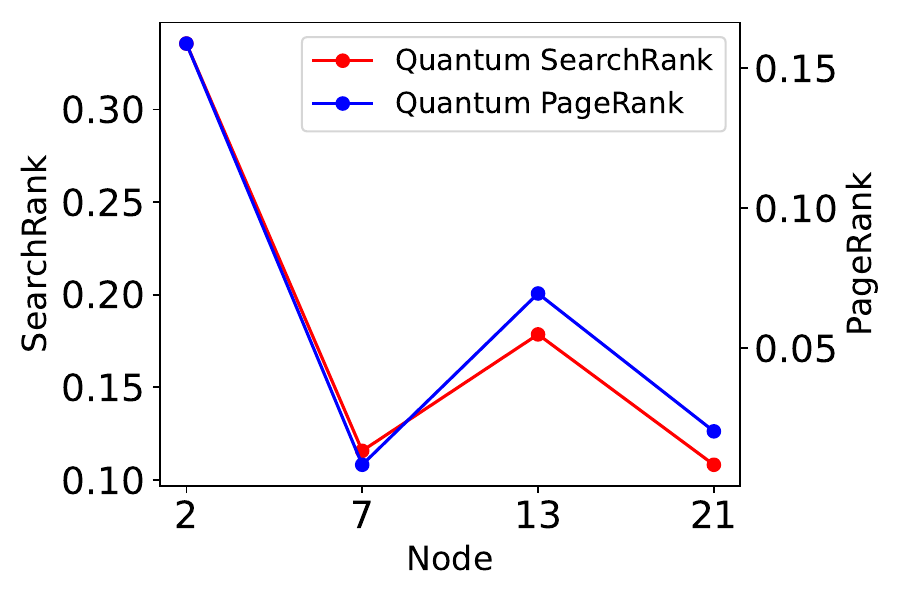}\label{F:classification_Q_Q}}
		\hspace{-6pt}
		\subfigure[]{\includegraphics[scale=0.45]{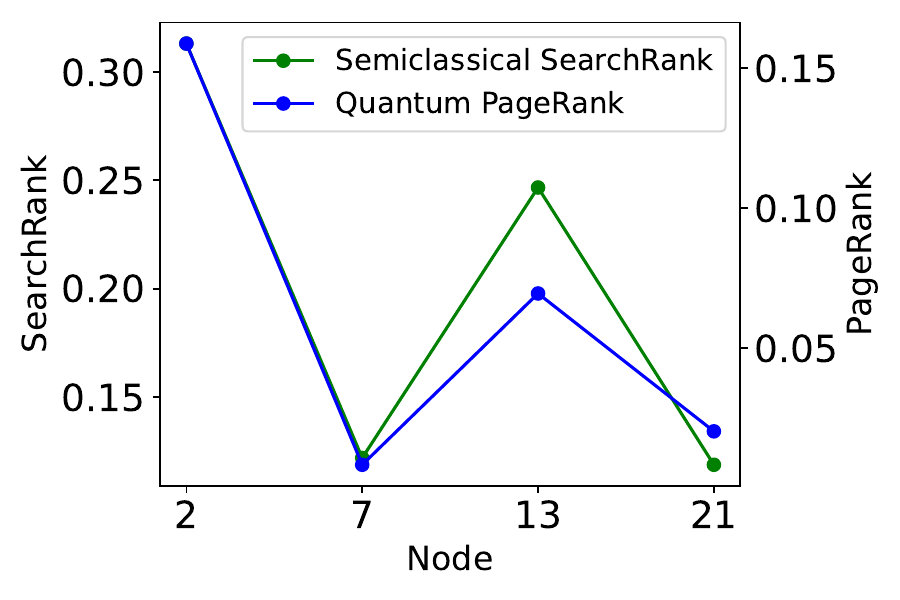}\label{F:classification_Q_S}}
		\hspace{-6pt}
		\subfigure[]{\includegraphics[scale=0.45]{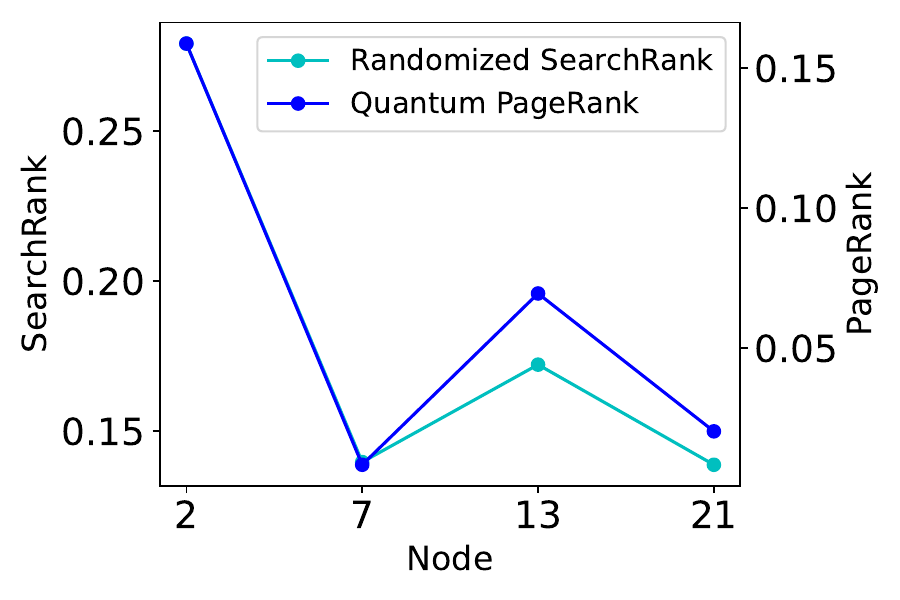}\label{F:classification_Q_Si}}
	}
	\caption{Comparison of the SearchRank distributions of the marked nodes with the PageRank distributions. a)-c) Comparison with the classical PageRank. d)-f) Comparison with the quantum PageRank. The PageRank (right axis) is represented on a different scale than the SearchRank (left axis).}
	\label{F:classification}
\end{figure*}

In the case of Szegedy's semiclassical walk, there are two classes of algorithms, class I and class II, depending on whether register 1 or 2 is being measured respectively to obtain the information about the position of the walker. Since in the quantum PageRank the second register is being measured, we will deal with the semiclassical walks of class II. Each classical step of the semiclassical walk consists on a quantum evolution determined by the quantum time $t_q$, a measurement of the position and a reset of the system. To reset the system, the information in the first register is erased, and the result of the position is used to prepare its corresponding proxy state. These states represent the classical position of the walker in the Hilbert space where the quantum walk takes place \cite{Semiclassical}. A priori these proxies could be defined arbitrarily. Nevertheless, in the case of Szegedy's semiclassical walk a natural choice for the proxy states is the set formed by the $\left|\psi_i\right>$ states in \eqref{psi_i}. The whole scheme of its implementation is shown in Figure \ref{F:class_II}.

Formally, the semiclassical walks can be represented as classical walks where the transition matrices encode the quantum evolution. These are called semiclassical transition matrices, and are defined as:
\begin{equation}\label{G2}
	\tensor[_2]{G}{}^{(t_q)}_{ji} := \left|\left|\tensor[_2]{\left<j\right|U^{t_q}\left|\psi_i\right>}{}\right|\right|^2,
\end{equation}
where the left subindex indicates that it is a semiclassical walk of class II.
Note that there is a different semiclassical walk for each value of the quantum time $t_q$. Thus, this quantity parametrizes a family of semiclassical walks.

\section{Semiclassical SearchRank algorithm}\label{Sc-SearchRank}

In this work we propose a semiclassical version of the quantum SearchRank algorithm, replacing the underlying Szegedy's quantum walk by a semiclassical walk.

If we substitute the general unitary $U$ by the SearchRank operator $W_Q$ in \eqref{G2}, we obtain the semiclassical transition matrices $\tensor[_2]{G}{}^{(t_q)}$ for the Semiclassical SearchRank algorithm. Each of these matrices can be treated as a Google matrix, so that their stationary distributions give us an instantaneous Semiclassical SearchRank distribution for each value of the quantum time $t_q$. Let us represent these distributions as column vectors and denote them as $S_{sc}(t_q)$, so that they satisfy the following matrix fixed-point equation:
\begin{equation}
	\tensor[_2]{G}{}^{(t_q)} S_{sc}(t_q) = S_{sc}(t_q).
\end{equation}

From an operational point of view, we have to perform each of these semiclassical walks over an initial probability distribution until they converge. An example is shown in Figure \ref{F:semiclassical} (upper panel). In this case, the initial distribution is the uniform one. Since the quantum circuit must be repeated to sample the final distribution, in order to initialize the classical distribution we prepare one of the $\left|\psi_i\right>$ states at random each repetition. Formally, this corresponds to the preparation of the mixed state
\begin{equation}\label{rho}
	\rho := \frac{1}{N} \sum_{i=0}^{N-1} \left|\psi_i\right>\left<\psi_i\right|.
\end{equation}
After the initialization, we perform the quantum evolution for the particular value of the quantum time $t_q$, measure the second register and reset for the following classical step. The process is repeated $t_c^*$ times, which is defined as the number of classical steps required for the semiclassical walk to converge. Finally, the outcome of the last measurement is used to sample the semiclassical SearchRank distribution.

Since the searching functionality of the algorithm is in the quantum evolution, there is a value of the quantum time $t_q$ for which the probability of measuring the marked nodes is maximum. As in the case of the quantum SearchRank, the time complexity will be close to $t_q \approx \sqrt{N/M}$, as will be proved in section \ref{time_complexity}. Thus, this algorithm maintains the same complexity with respect to the evolution of quantum time. Nevertheless, the semiclassical walk requires to repeat the quantum evolution the number of classical steps $t_c^*$ required to converge. Thus, the actual number of times that the operator $W_Q$ is called is $t_q \times t_c^*$.

In principle, the value of $t_c^*$ may depend on the size $N$ of the graph, as in the classical PageRank algorithm \cite{Adiabatic_PR}, worsening the time complexity of the semiclassical algorithm with respect to the quantum one. In order to overcome this issue, we propose a simplification fixing the number of classical steps. In particular, we will analyze the extreme case where only one classical step is carried out. This simplification is shown in Figure \ref{F:semiclassical} as a blue dashed box and we shall refer to it as the Randomized SearchRank algorithm.

As we have mentioned above, the initialization of the semiclassical walk in a quantum computer is equivalent to the preparation of the density matrix \eqref{rho}. Thus, in the extreme case where only one classical step is performed, the algorithm is equivalent to a quantum walk over the randomized mixed state instead of the equal superposition \eqref{initial}. A more rigorous proof can be found in \cite{Squwals}. Therefore, the quantum circuit scheme is very similar to that of the quantum SearchRank shown in Figure \ref{F:semiclassical} (bottom left panel). In both algorithms the quantum time corresponds directly to the total time of the walk.

Although this last algorithm is the one that we are interested on in this work, we will also analyze the performance of the full semiclassical algorithm for comparison.

\subsection{Numerical Simulations}

To study the Semiclassical SearchRank algorithm we need to use a classical simulator, since fault-tolerant quantum computers are not yet available. Until very recently, classical simulators have needed time and memory resources scaling as $\mathcal{O}(N^3)$ for dense transition matrices as the Google matrix. Moreover, a semiclassical walk requires to simulate the quantum walk over all the $N$ $\left|\psi_i\right>$ states, so that the time required was $\mathcal{O}(N^4)$.

Nevertheless, recently a new algorithm that saves both time and memory resources has been devised. This algorithm is implemented in the python library SQUWALS \cite{Squwals}. It only requires time and memory resources scaling as $\mathcal{O}(N^2)$ for a single Szegedy's quantum walk, so that the total time required to simulate the Semiclassical SearchRank algorithm scales as $\mathcal{O}(N^3)$. We have used this library in all the simulations performed in this work.

\begin{figure*}
	\makebox[0pt][c]{
		\subfigure[]{\includegraphics[scale=0.55]{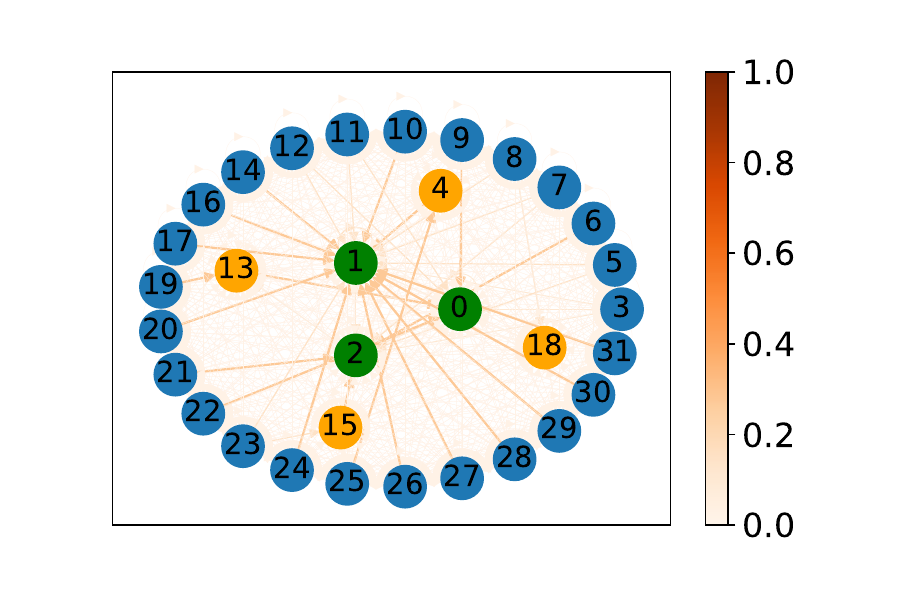}\label{F:weighted_google}}
		\subfigure[]{\includegraphics[scale=0.55]{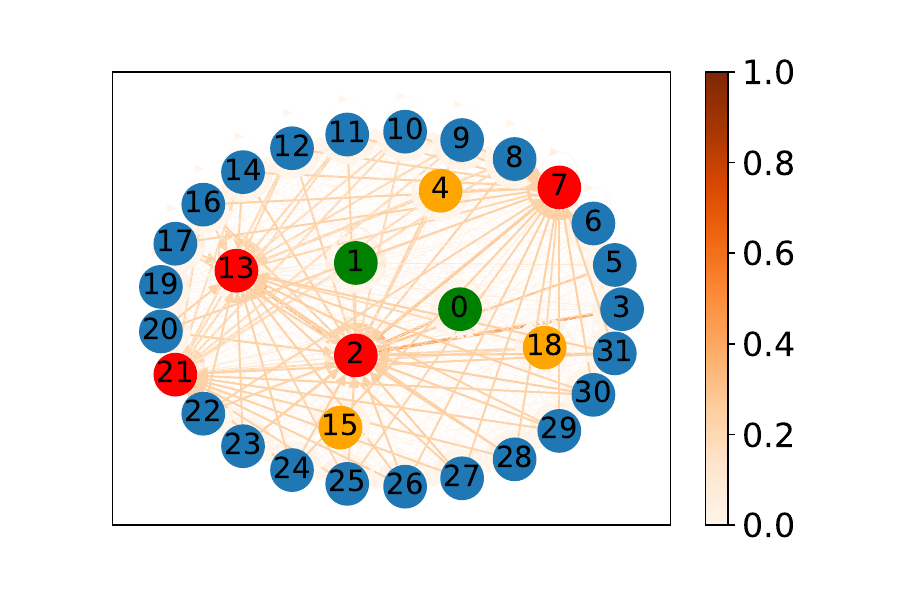}\label{F:weighted_semiclassical}}
	}
	\\
	\makebox[0pt][c]{
		\hspace{-24pt}
		\subfigure[]{\includegraphics[scale=0.55]{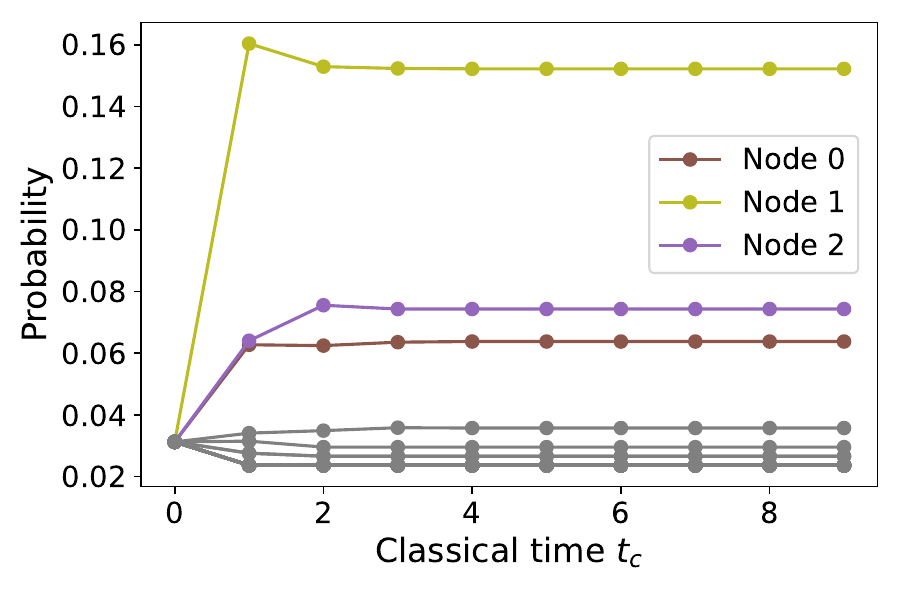}\label{F:weighted_convergence}}
		\subfigure[]{\includegraphics[scale=0.55]{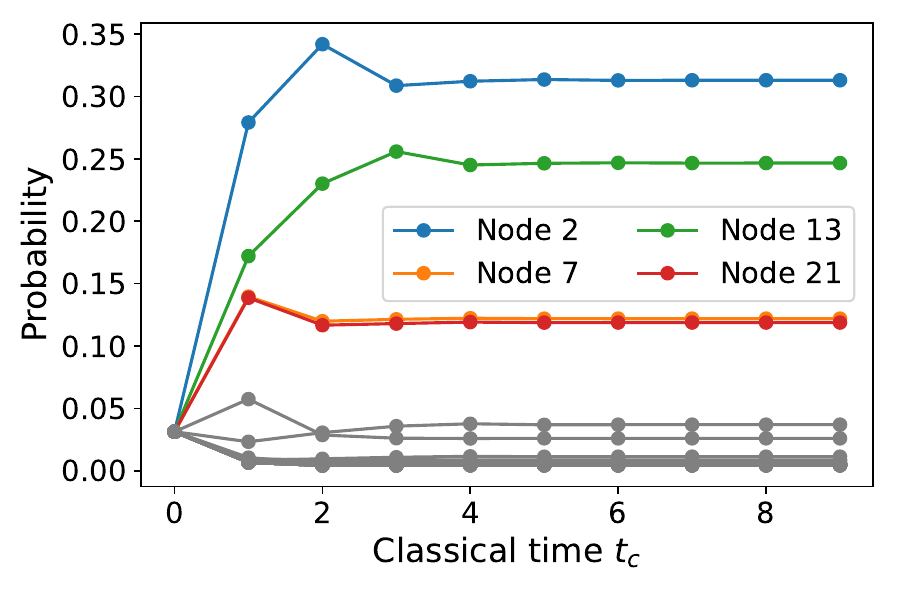}\label{F:weighted_convergence_semiclassical}}
	}
	\caption{a) Weighted graph representing the Google matrix with $\alpha = 0.25$ for the network with 32 nodes of Figure \ref{F:graph}. The strongest links point to the main nodes in green. b) Weighted graph representing the semiclassical matrix of the Semiclassical SearchRank for $t_q = 2$. In this case the strongest links point to the marked nodes in red. c) Probability of each node versus the classical time for the classical walk represented in a). The probability attains a higher value for the main nodes. d) Probability of each node versus the classical time for the semiclassical walk represented in b). The probability of the marked nodes is amplified whereas the other nodes obtain a residual SearchRank.}
	\label{F:weighted_graphs}
\end{figure*}

\section{Example on a scale-free graph}\label{Example}

In this section we will look at one example of the SearchRank algorithms using a scale-free graph. This type of graphs are not only good models for the World Wide Web \cite{SF-WWW}, but also have a wide range of applications such as in neural networks \cite{SF-Brain}, metabolomics \cite{SF-Metabolism-1,SF-Metabolism-2} and finance \cite{SF-Finances}.

We have created a random scale-free network using the python library NetworkX \cite{NetworkX} with the default parameters. This graph, shown in Figure \ref{F:graph}, was previously studied in \cite{APR} in the context of quantum PageRank. Due to the way the network is constructed \cite{Directed-SF}, the first nodes (inner green) have the most links pointing to them. Therefore, they are expected to be the most important in the classical PageRank. The middle orange nodes are secondary nodes that have few internal links. Finally, the outer blue nodes are residual nodes, which lack links pointing to them, and will have minimal degenerate classical PageRank. As for the quantum PageRank, this algorithm breaks the degeneracy of the residual nodes, so that some of them may receive higher importance than the secondary nodes \cite{Paparo2,APR}. The classical and quantum PageRank distributions are shown in Figure \ref{F:histogram}.

For the SearchRank algorithms we have marked four nodes, namely, node 2, which is one of the most important, the secondary node 13 and the residual nodes 7 and 21. The probability of finding one of these nodes at each value of the quantum time is shown in Figure \ref{F:probabilities} for the three SearchRank algorithms we have considered in this work. In this case it is maximum at $t_q = 3$ for the Randomized SearchRank, while it is maximum at $t_q = 2$ for the quantum and semiclassical algorithms. In all three cases the probability is greater than $0.7$, so there are many possibilities to measure them. However, this is only one example. In a later section we will analyze the probability achieved by the SearchRank algorithms for graphs of increasing size and different number of marked nodes, as well as the quantum time complexity of the algorithms.

As we can see in Figure \ref{F:histogram}, the four marked nodes have the greatest ranking in the three SearchRank distributions, so that their probability has been effectively amplified. In order to compare the ranking of the marked nodes with respect to the PageRank distributions, we have isolated their distributions in Figure \ref{F:classification}. Since the probability of the marked nodes has been amplified in the SearchRank algorithms, we represent the SearchRank distributions on a different scale than the PageRank distributions. The most important of the marked nodes, both in the classical and quantum PageRank, is node 2, and this one has been properly detected as the most important node by the three SearchRank algorithms. The second node in importance is node 13, which is a secondary node of the network. Again, the three SearchRank algorithms detect it properly as the second most important node. So far the order of importance is maintained. However, at nodes 7 and 21 there is a violation of the order. In all three SearchRank algorithms, node 7 is more important than node 21, while in the classical PageRank they are degenerate. Nevertheless, in the case of the Randomized SearchRank they are almost degenerate, as can be seen in Figure \ref{F:classification_C_Si}. In the quantum PageRank, node 21 is more important than node 7, so the order is reversed.

\begin{figure*}
	\makebox[0pt][c]{
		\subfigure[]{\includegraphics[scale=0.45]{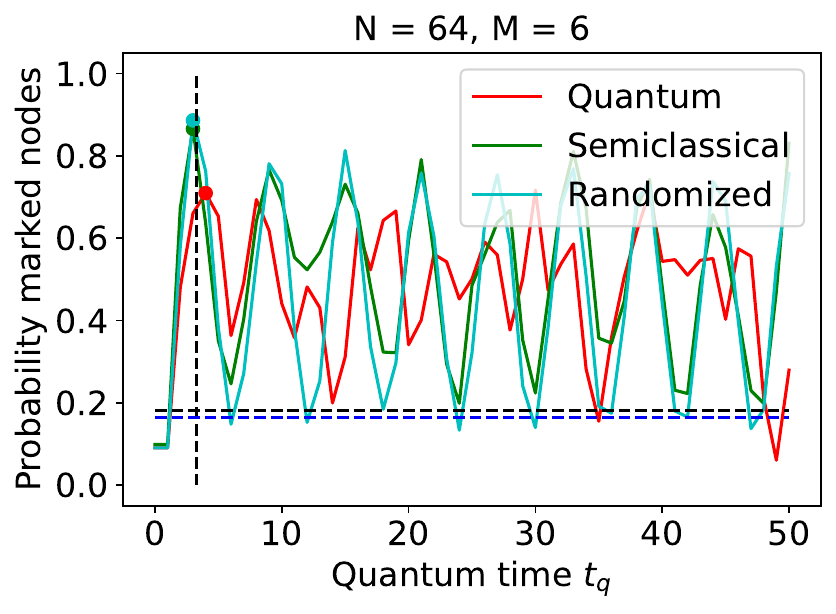}\label{F:probability_64_6}}
		\hspace{-6pt}
		\subfigure[]{\includegraphics[scale=0.45]{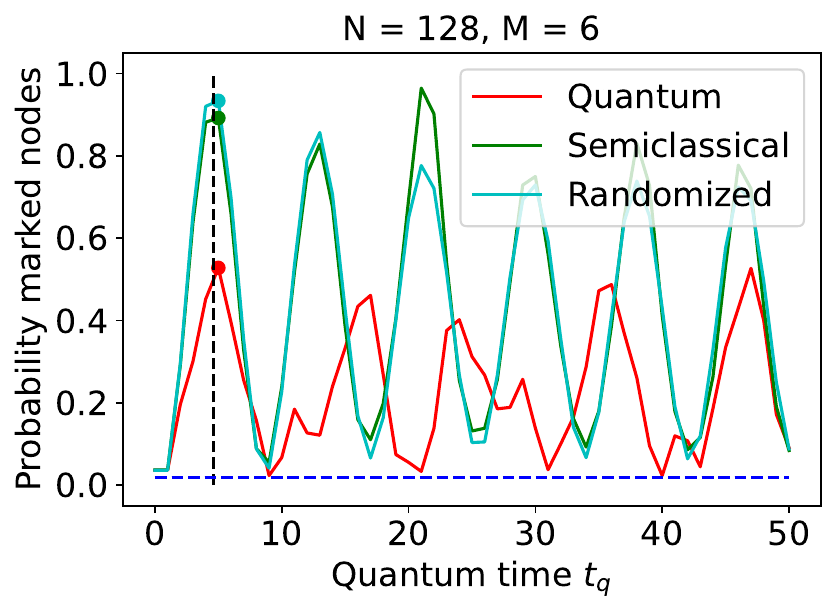}\label{F:probability_128_6}}
		\hspace{-6pt}
		\subfigure[]{\includegraphics[scale=0.45]{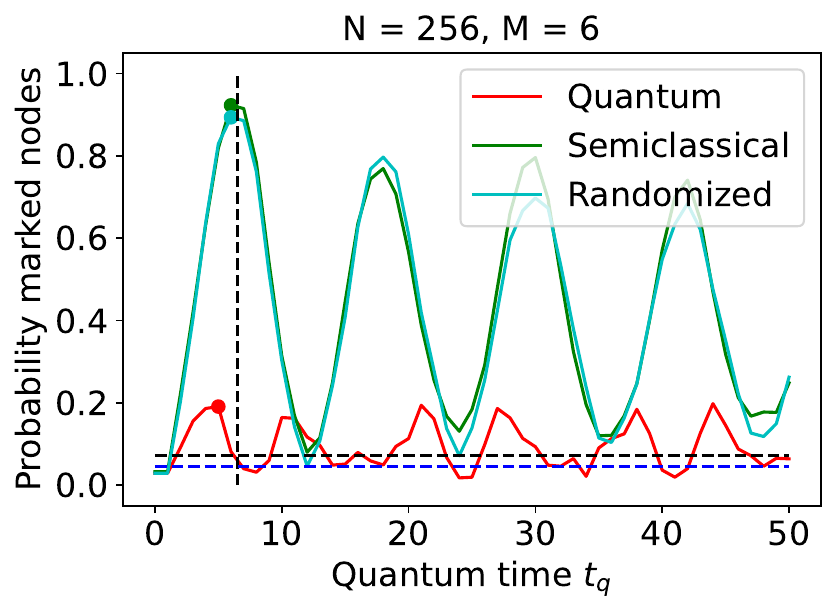}\label{F:probability_256_6}}
	}
	\\
	\makebox[0pt][c]{
		\subfigure[]{\includegraphics[scale=0.45]{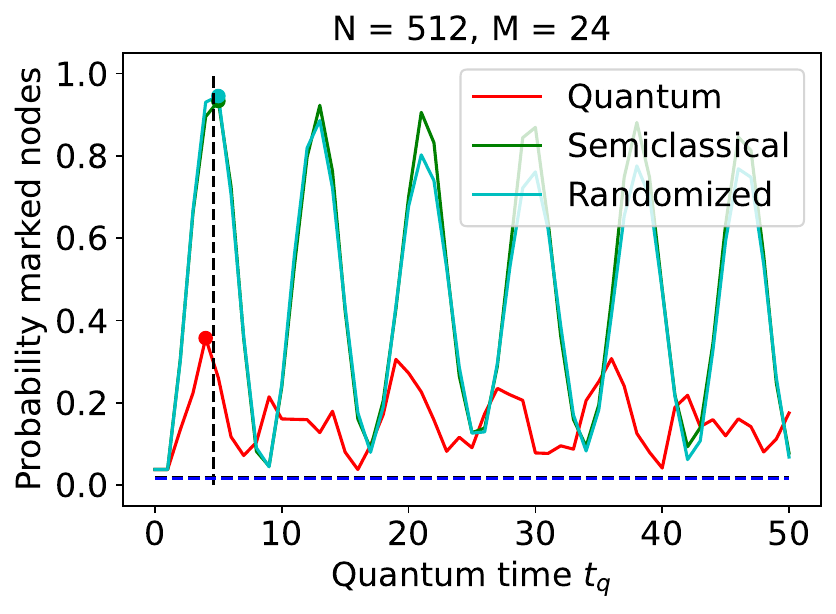}\label{F:probability_512_24}}
		\hspace{-6pt}
		\subfigure[]{\includegraphics[scale=0.45]{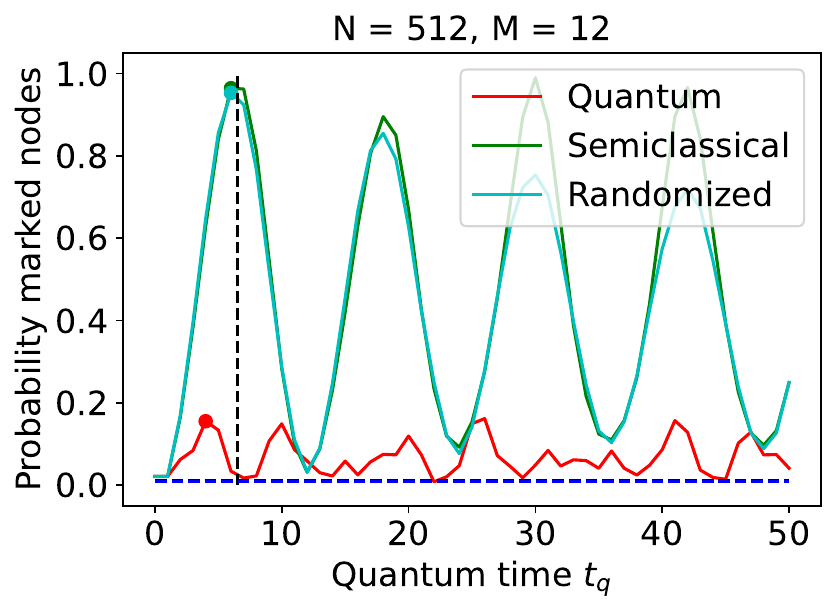}\label{F:probability_512_12}}
		\hspace{-6pt}
		\subfigure[]{\includegraphics[scale=0.45]{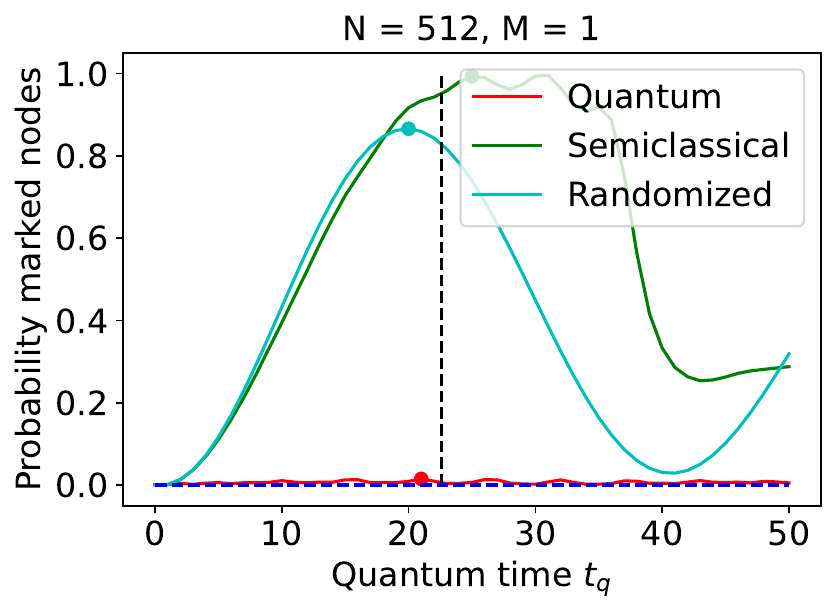}\label{F:probability_512_1}}
	}
	\caption{Probability of measuring one of the marked nodes versus the quantum time for the three SearchRank algorithms in different scale-free graphs. a) In a graph with $N=64$ nodes and $M=6$ marked nodes. b) In a graph with $N=128$ nodes and $M=6$ marked nodes. c) In a graph with $N=256$ nodes and $M=6$ marked nodes. d) In a graph with $N=512$ nodes and $M=24$ marked nodes. e) In a graph with $N=512$ nodes and $M=12$ marked nodes. f) In a graph with $N=512$ nodes and $M=1$ marked node. The first maximum of each curve is marked with a dot. The vertical dashed line represents the reference time $\sqrt{N/M}$. The horizontal dashed lines represent the probability of the marked nodes in the classical (black) and quantum (blue) PageRank distributions.}
	\label{...}
\end{figure*}

In \cite{APR} it was shown that the quantum PageRank can introduce fluctuations in the classically degenerate node order that can be misleading. Thus, the SearchRank algorithms are expected to introduce fluctuations that may be different, and it is not uncommon for the residual node order to be reversed with respect to the quantum PageRank. Therefore, in general it seems that all three SearchRank algorithms correctly rank the nodes. In a later section we will study statistically the agreement between the PageRank and SearchRank distributions for a large set of different scale-free networks and marked nodes.

\subsection{Visualizing the Semiclassical Search}

As mentioned above, a semiclassical walk is like a classical walk but with a semiclassical transition matrix encoding the quantum evolution. Thus, we can use the semiclassical transition matrix to represent the semiclassical walk as a weighted graph. This will allow us to visualize how the probability of the marked nodes in the search process is amplified.

In Figure \ref{F:weighted_google} we have represented the weighted network whose weights are given by the Google matrix \eqref{G} using $\alpha = 0.25$. As expected, we can observe a large flow of information to the main nodes of the graph, and some flow to the secondary nodes. Thus, this classical walk gives a higher rank to these nodes in the limiting distribution, as shown in Figure \ref{F:weighted_convergence}, where the probability of each node at each classical step is plotted. When we perform the semiclassical walk from this graph with $t_q=2$, it is equivalent to a classical walk whose weighted graph is shown in Figure \ref{F:weighted_semiclassical}. We can now observe that the flow of information is mainly directed to the marked nodes (red) and, therefore, there will be a high probability of measuring them in the limiting distribution. As shown in Figure \ref{F:weighted_convergence_semiclassical}, the probability of the marked nodes converges rapidly to an amplified value, while the rest of the nodes adopt a residual classification. Therefore, the semiclassical algorithm modifies the classical network to search for the marked nodes.

\section{Searching Power of the SearchRank Algorithms}\label{Searching}

\begin{table*}[htbp]
	\centering
	\caption{Parameters of the scaling law \eqref{fit_function} for the fits to the measurement probability data of the marked nodes and quantum time complexity.}
	\begin{tabular}{ccccc}
		\toprule
		Function & Parameter & Quantum & Semiclassical & Randomized \\
		\midrule
		\multicolumn{1}{c}{\multirow{2}[2]{*}{Optimal time}} & n     & $0.455 \pm 0.016$ & $0.523 \pm 0.009$ & $0.473 \pm 0.008$ \\
		& A     & $1.19 \pm 0.08$ & $0.91 \pm 0.03$ & $1.05 \pm 0.03$ \\
		\midrule
		\multicolumn{1}{c}{\multirow{2}[2]{*}{Optimal probability}} & n     & $-1.046 \pm 0.035$ & ---   & --- \\
		& A     & $11.45 \pm 1.90$ & ---   & --- \\
		\midrule
		\multicolumn{1}{c}{\multirow{2}[2]{*}{Probability reference time}} & n     & $-1.109 \pm 0.078$ & ---   & --- \\
		& A     & $6.20 \pm 2.29$ & ---   & --- \\
		\bottomrule
	\end{tabular}%
	\label{T:fittings}%
\end{table*}%

One of the two virtues of the SearchRank algorithms is their ability to amplify the probability of measuring the marked nodes. In this section we will analyze what is the probability that each algorithm can achieve, and what is the time complexity to reach the maximum probability.

\subsection{Probability at the Maximum}

\begin{figure*}
	\subfigure[]{\includegraphics[scale=0.5]{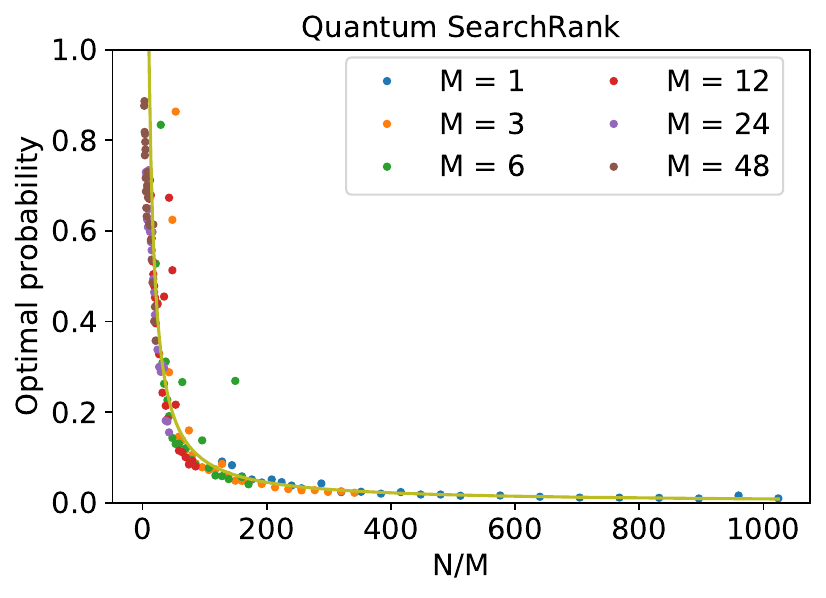}\label{F:maxima_probability_quantum}}
	\subfigure[]{\includegraphics[scale=0.5]{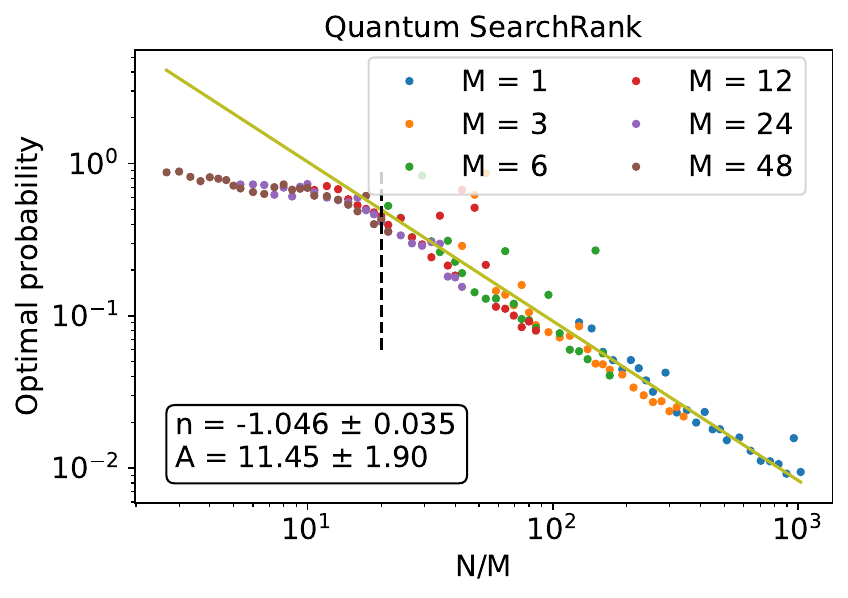}\label{F:maxima_probability_quantum_log}}
	\subfigure[]{\includegraphics[scale=0.5]{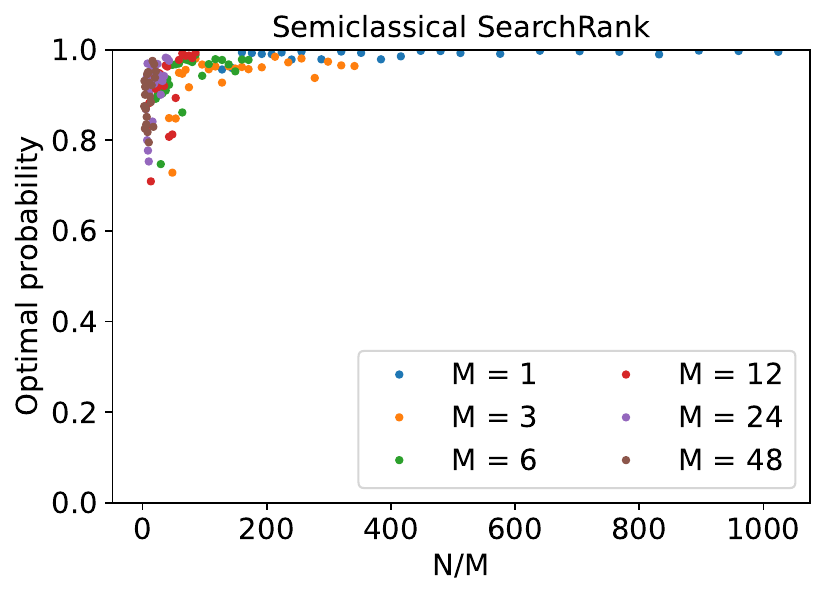}\label{F:maxima_probability_semiclassical}}
	\subfigure[]{\includegraphics[scale=0.5]{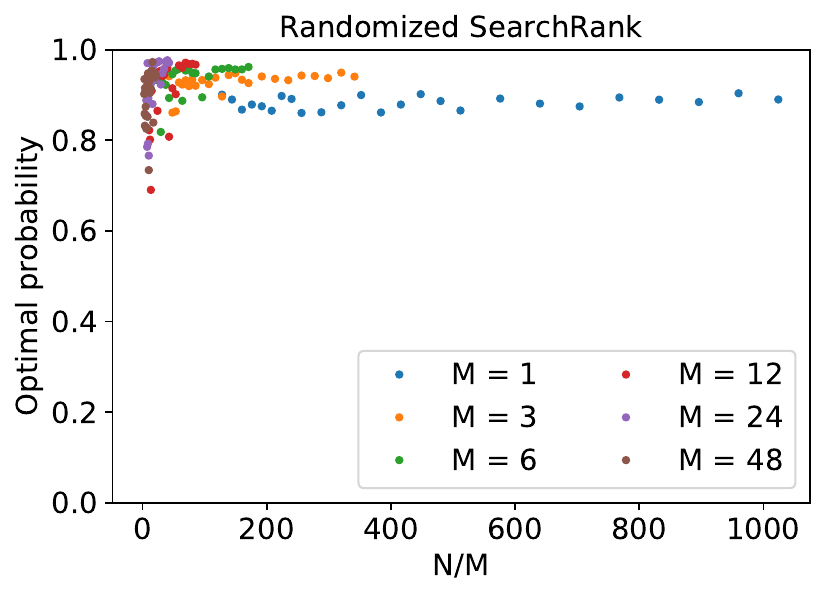}\label{F:maxima_probability_simplified}}
	\caption{Probability at the maximum versus the relation $N/M$ for a) the quantum SearchRank, b) the quantum SearchRank in logarithmic scale, c) the semiclassical SearchRank and d) the Randomized SearchRank. The vertical dashed line in b) indicates the region from which the linear fit has been done.}
	\label{F:maxima_probability}
\end{figure*}

In the previous example, all SearchRank algorithms were able to extend the probability of the marked nodes above $0.7$. However, in this section we will show that the probability at the maximum drops with $N/M$ for the quantum algorithm, while it remains at a high value for the Semiclassical and Randomized SearchRank algorithms. Although here we will deal with the actual maximum of the probability curves, in a later section we will show that it also remains at a high value for the reference measurement time $t_q = \sqrt{N/M}$.

On the one hand, to show that the quantum probability drops with the network size $N$, we simulated the SearchRank algorithms for scale-free random graphs with $N = 64$, $128$ and $256$ nodes, all three with $M = 6$ randomly chosen marked nodes. The probability curves with respect to quantum time are shown in Figures \ref{F:probability_64_6}-\ref{F:probability_256_6}. We can observe how indeed the maximum point of the curve corresponding to the quantum algorithm reaches a lower value as the size of the graph $N$ increases. On the other hand, in order to analyze the effect of the number of marked nodes, we have simulated the algorithms for three scale-free random graphs with $N = 512$ nodes, and $M = 24$, $12$ and $1$ marked nodes. The corresponding probability curves are shown in Figures \ref{F:probability_512_24}-\ref{F:probability_512_1}. In this case, the probability of measuring a marked node in the quantum SearchRank decreases as the number $M$ of nodes decreases.

Now that we have demonstrated the qualitative relationship between the quantum probability and the $N$ and $M$ parameters of the problem, let us obtain a quantitative relationship. To do so, we have simulated the SearchRank algorithms for different scale-free random networks with sizes ranging from $N = 64$ to $N = 1024$. For all networks we have chosen $M = 1$, $3$, $6$, $12$, $24$ and $48$ nodes at random. Since some results of quantum search problems, such as Grover's algorithm, depend directly on the $N/M$ ratio \cite{Portugal}, we have plotted the maximum likelihoods with respect to $N/M$ in Figure \ref{F:maxima_probability}.

In the case of the quantum SearchRank, in Figure \ref{F:maxima_probability_quantum} we can observe that the probability drops rapidly with $N/M$ as expected. To obtain a mathematical expression, we intend to fit the data to the following scaling law:
\begin{equation}\label{fit_function}
	f(N/M) = A\left(\frac{N}{M}\right)^n.
\end{equation}
We have plotted the same data in logarithmic scale in Figure \ref{F:maxima_probability_quantum_log}. Although there is an initial region where the probability remains very high, from $N/M \approx 20$ there is a clear linear relationship. A linear fit in this asymptotic region yields an exponent $n = -1.046$, so that the quantum probability falls asymptotically as approximately $\mathcal{O}(M/N)$. The parameters of the fit are summarized in Table \ref{T:fittings}.

\begin{figure*}
	\makebox[0pt][c]{
		\subfigure[]{\includegraphics[scale=0.45]{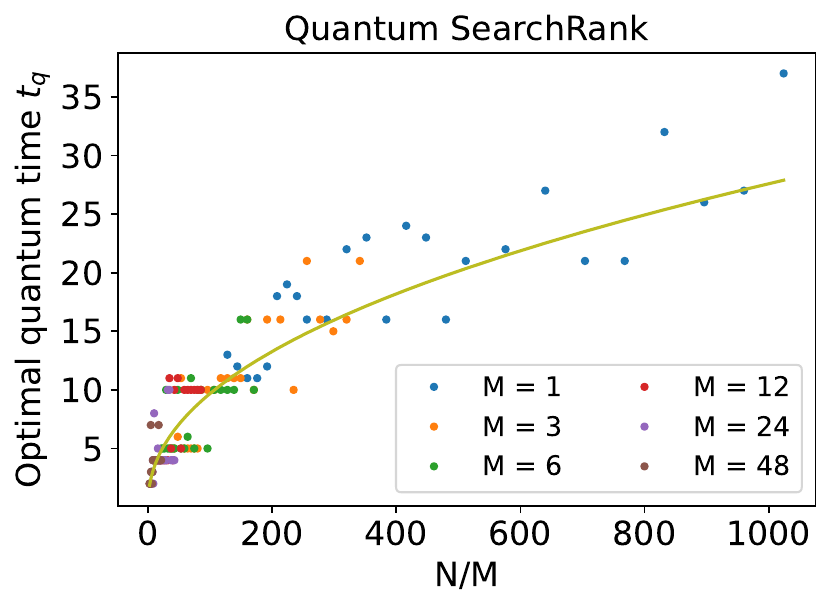}\label{F:maxima_time_quantum}}
		\hspace{-6pt}
		\subfigure[]{\includegraphics[scale=0.45]{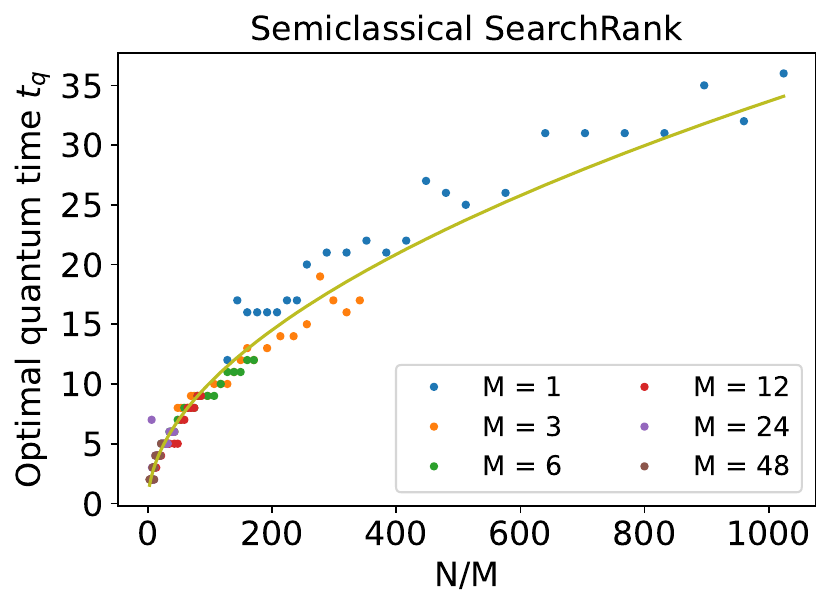}\label{F:maxima_time_semiclassical}}
		\hspace{-6pt}
		\subfigure[]{\includegraphics[scale=0.45]{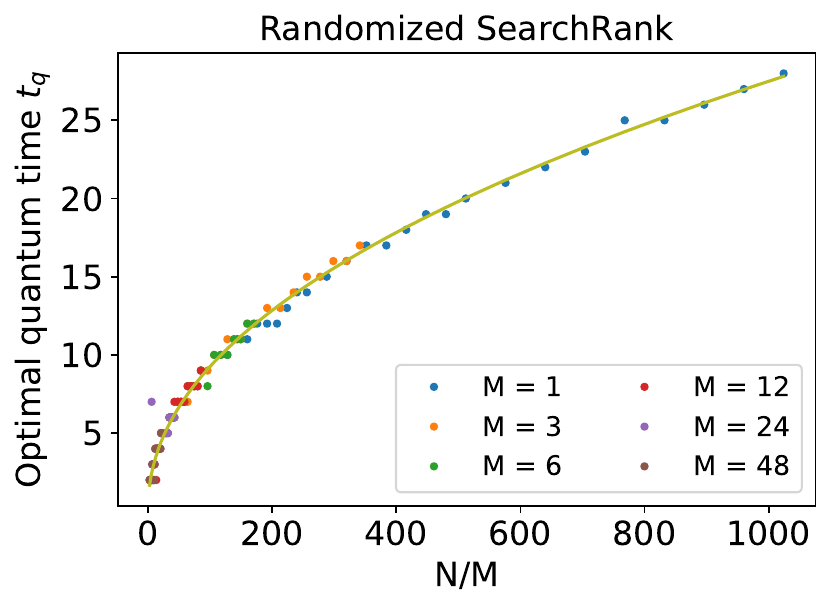}\label{F:maxima_time_simplified}}
	}
	\\
	\makebox[0pt][c]{
		\subfigure[]{\includegraphics[scale=0.45]{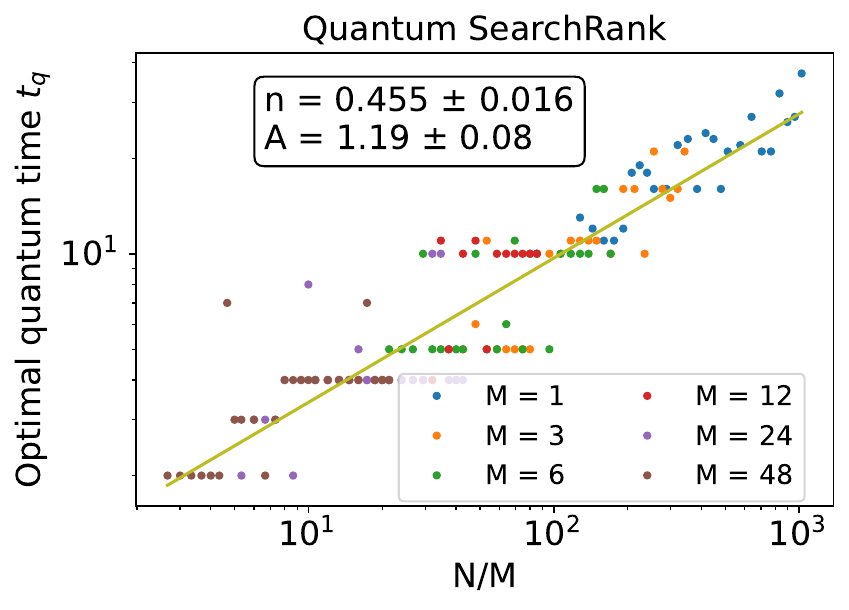}\label{F:maxima_time_quantum_log}}
		\hspace{-6pt}
		\subfigure[]{\includegraphics[scale=0.45]{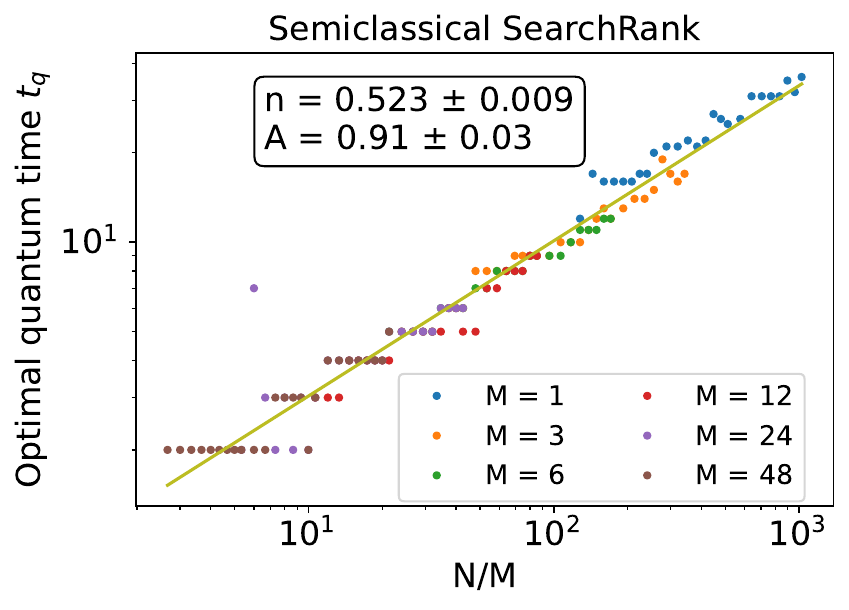}\label{F:maxima_time_semiclassical_log}}
		\hspace{-6pt}
		\subfigure[]{\includegraphics[scale=0.45]{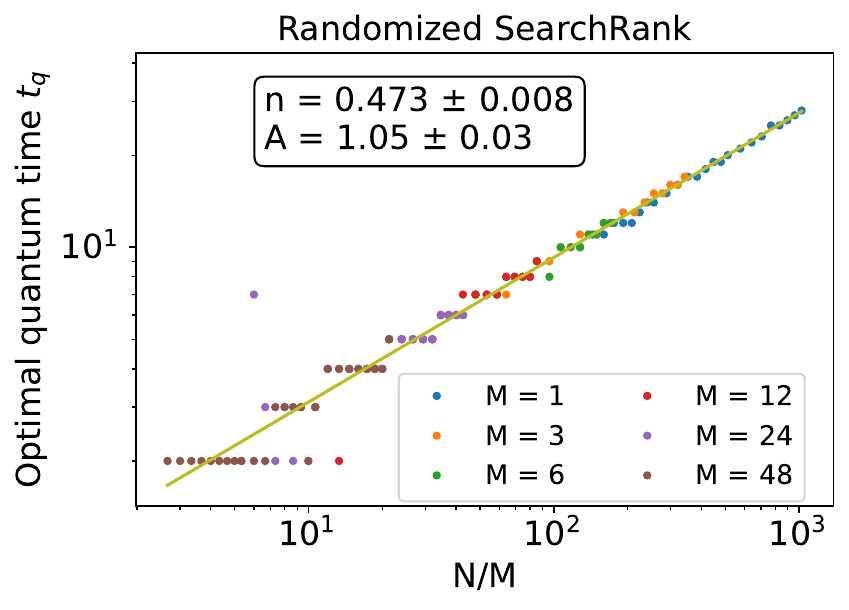}\label{F:maxima_time_simplified_log}}
	}
	\caption{Quantum time for which the maximum of probability occurs versus the relation $N/M$ for a) the quantum SearchRank, b) the Semiclassical SearchRank and c) the Randomized SearchRank. The corresponding data for the linear fit is represented in d)-f) in logarithmic scale.}
	\label{F:maxima_time}
\end{figure*}

For the Semiclassical SearchRank algorithm, in Figure \ref{F:maxima_probability_semiclassical} we can see that in the asymptotic region the maximum probability is close to $1$. Finally, in Figure \ref{F:maxima_probability_simplified} the maximum probability for the Randomized SearchRank is represented. In this case the probability is a little lower, but also close to $1$. Table \ref{T:probabilities} shows the average probability achieved in the asymptotic region. It seems that the results do not depend on the number of marked nodes $M$, as expected, and the probability remains around $0.9$. We can conclude, therefore, that the semiclassical approach solves the problem with the probability of the quantum SearchRank algorithm, even after simplification.

\begin{table}[htbp]
	\centering
	\caption{Average probability of measuring one of the marked nodes in the asymptotic region for the Semiclassical and Randomized SearchRank algorithms and different number M of marked nodes. Both measuring at the optimal time and at the reference time, showing that a high probability around $0.9$ is obtained.}
	\begin{tabular}{c|cc|cc}
		\hline
		\multirow{3}[4]{*}{M} & \multicolumn{2}{c|}{Optimal} & \multicolumn{2}{c}{Reference time} \\
		\cline{2-5}          & \multirow{2}[2]{*}{Semiclassical} & \multirow{2}[2]{*}{Randomized} & \multirow{2}[2]{*}{Semiclassical} & \multirow{2}[2]{*}{Randomized} \\
		&       &       &       &  \\
		\hline
		1     & $0.99 \pm 0.00$ & $0.89 \pm 0.01$ & $0.97 \pm 0.01$ & $0.85 \pm 0.02$ \\
		3     & $0.97 \pm 0.01$ & $0.94 \pm 0.01$ & $0.94 \pm 0.05$ & $0.91 \pm 0.01$ \\
		6     & $0.97 \pm 0.01$ & $0.95 \pm 0.02$ & $0.95 \pm 0.03$ & $0.94 \pm 0.02$ \\
		12    & $0.94 \pm 0.07$ & $0.94 \pm 0.05$ & $0.93 \pm 0.07$ & $0.93 \pm 0.05$ \\
		24    & $0.95 \pm 0.03$ & $0.96 \pm 0.02$ & $0.95 \pm 0.03$ & $0.96 \pm 0.02$ \\
		48    & $0.92 \pm 0.04$ & $0.93 \pm 0.04$ & $0.92 \pm 0.04$ & $0.93 \pm 0.04$ \\
		\hline
	\end{tabular}%
	\label{T:probabilities}%
\end{table}%

\subsection{Time Complexity}\label{time_complexity}

Despite the fact that we have shown how the Semiclassical SearchRank is able to amplify effectively the probability of measuring the marked nodes, the question that arises is whether the time complexity is similar to that of the quantum algorithm. In this section we will examine the quantum time scaling for which the maximum probability occurs.

Again, we expect the time value of the maximum to depend on the $N/M$ ratio. Thus, we have plotted these time values with respect to $N/M$ in Figures \ref{F:maxima_time_quantum}-\ref{F:maxima_time_simplified} for the three SearchRank algorithms. As expected, there is a clear relationship with the $N/M$ quantity. To fit the data to the scaling \eqref{fit_function}, we have plotted the same data in logarithmic scale in Figures \ref{F:maxima_time_quantum_log}-\ref{F:maxima_time_simplified_log}. The results of those fits are summarized in Table \ref{T:fittings}. The linear fits give an exponent of $n = 0.455$ for the quantum SearchRank, $n = 0.523$ for the Semiclassical SearchRank and $n = 0.473$ for the Randomized SearchRank. As for the prefactor $A$, we obtain $A = 1.19$ for the quantum SearchRank, $A = 0.91$ for the Semiclassical SearchRank and $A = 1.05$ for the Randomized SearchRank. Thus, we have found that the optimal measurement point in the semiclassical algorithm and its simplification is close to $t_q = \sqrt{N/M}$. In the case of the quantum SearchRank, it seems to be slightly faster than the others. Nonetheless, it should be noted that the data fluctuates a lot. This is mainly due to the difficulty of identifying the maximum when the probability is very small. Therefore, we can conclude that the semiclassical approach does not worsen the time complexity of the quantum algorithm, being relatively the same.

\subsection{Reference Time of Measurement}

\begin{figure*}
	\subfigure[]{\includegraphics[scale=0.5]{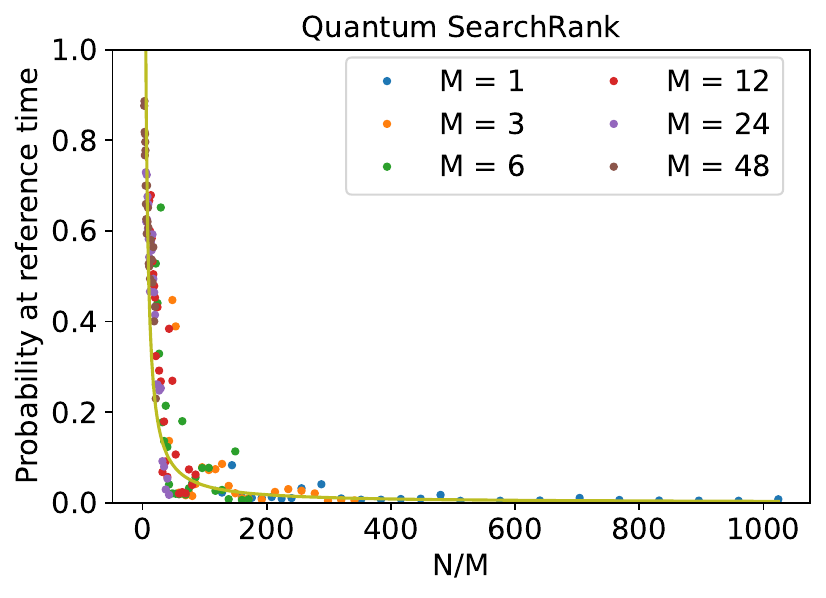}\label{F:reference_probability_quantum}}
	\subfigure[]{\includegraphics[scale=0.5]{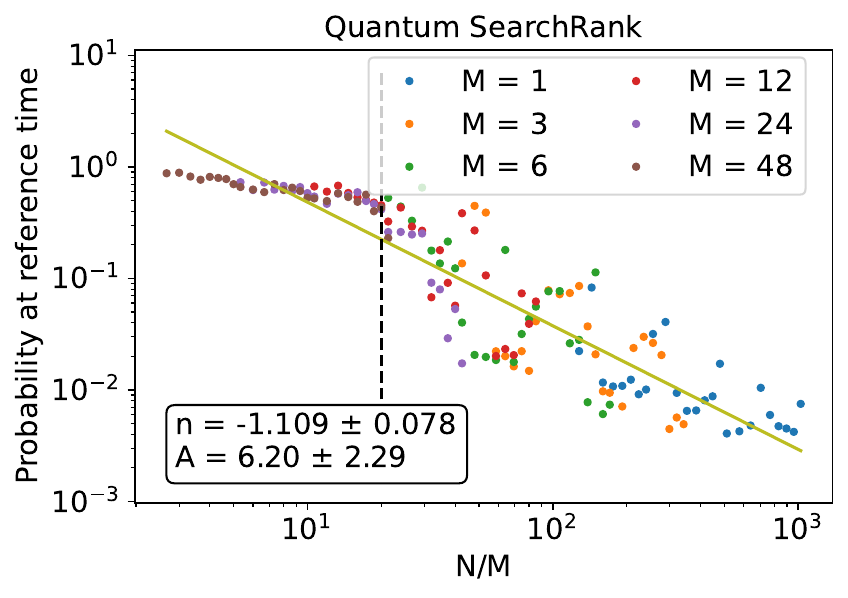}\label{F:reference_probability_quantum_log}}
	\subfigure[]{\includegraphics[scale=0.5]{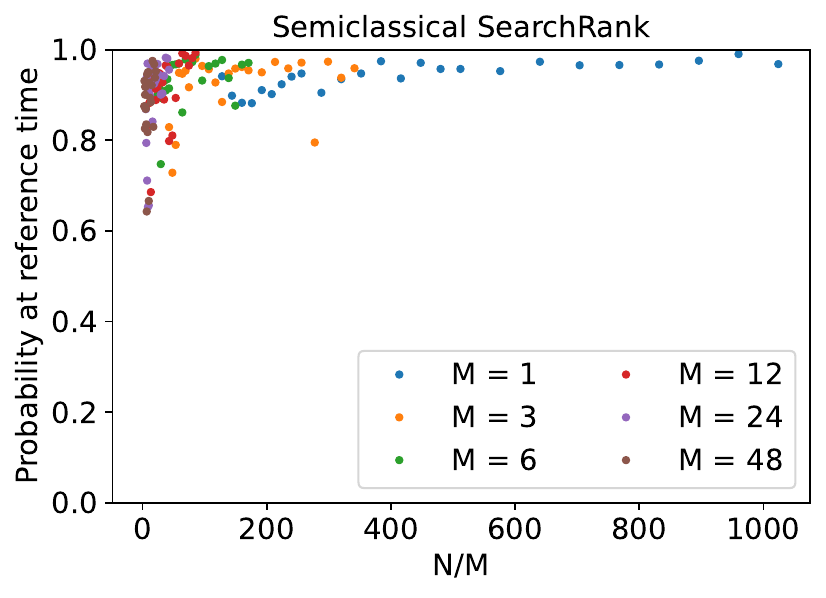}\label{F:reference_probability_semiclassical}}
	\subfigure[]{\includegraphics[scale=0.5]{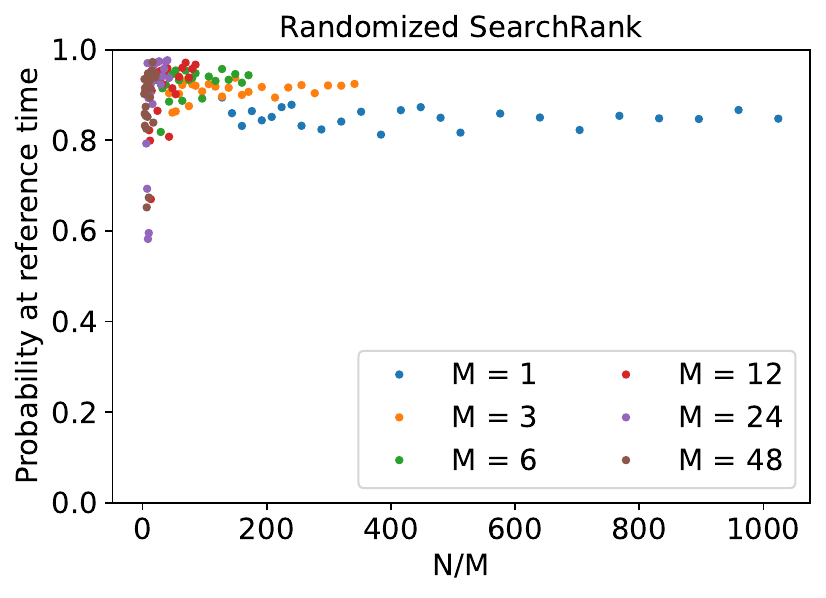}\label{F:reference_probability_simplified}}
	\caption{Probability at the reference time of measurement versus the relation $N/M$ for a) the quantum SearchRank, b) the quantum SearchRank in logarithmic scale, c) the Semiclassical SearchRank and d) the Randomized SearchRank. The vertical dashed line in b) indicates the region from which the linear fit has been done.}
	\label{F:reference_probability}
\end{figure*}

So far, we have observed that the Semiclassical and Randomized SearchRank algorithms are able to find the marked nodes with a high probability at maxima. However, to measure with this probability, we would need to know a priori how many quantum steps are needed to reach the maximum. Although we have seen that the optimal value of the quantum time is close to $\sqrt{N/M}$, we cannot be sure where the maximum is since it depends on the particular network.

As a reference, we can always measure to the nearest integer $t_q = \left\lfloor\sqrt{N/M}\right\rceil$, expecting to be close to the optimal point. Therefore, we need to analyze what is the real probability when measuring at this reference time. For this purpose, we have plotted in Figure \ref{F:reference_probability} the probability of measuring the marked nodes at the reference time. The probability in the quantum algorithm falls as expected, with a relationship with $N/M$ similar to the previous one. The parameters of the fit are summarized in Table \ref{T:fittings}. For the Semiclassical and Randomized algorithms the probability is very similar to the maximum. Therefore, we still measure the marked nodes with a very high probability. The average probability achieved by the Semiclassical and Randomized SearchRank algorithms is summarized in Table \ref{T:probabilities}.

We have shown that the Semiclassical SearchRank is an algorithm capable of performing effective quantum search on scale-free networks that the quantum algorithm does not perform. With simplification as a randomized quantum walk, we do not have to worry about the classical time to convergence, thereby the time complexity is similar to that of the quantum algorithm. Thus, we have devised an algorithm that can be useful for searching problems with a quadratic speedup with respect to classical algorithms, regardless of their ability to classify nodes.

\section{Ranking Power of the SearchRank Algorithms}\label{Ranking}

The second feature of the SearchRank algorithms is their ability to rank the marked nodes according to their importance. Since the Randomized SearchRank is a type of Semiclassical SearchRank, the natural question arises as to which algorithm it resembles the most in terms of ranking, the classical or the quantum PageRank. In other words, whether it is more classical or more quantum with respect to this property. In this section we are going to compare the ranking provided by the three SearchRank algorithms with the ranking given by both the classical and quantum PageRank. For this purpose we are going to use Kendall's coefficient \cite{Kendall}. It is used to measure the similarity between two ordered lists of items. Kendall's coefficient will be $1$ if both lists are equal, $-1$ if the order is totally reversed, $0$ if there is a total absence of correlation and intermediate values depending on the partial correlation.

\begin{figure*}[htpb]
	\subfigure[]{\includegraphics[scale=0.5]{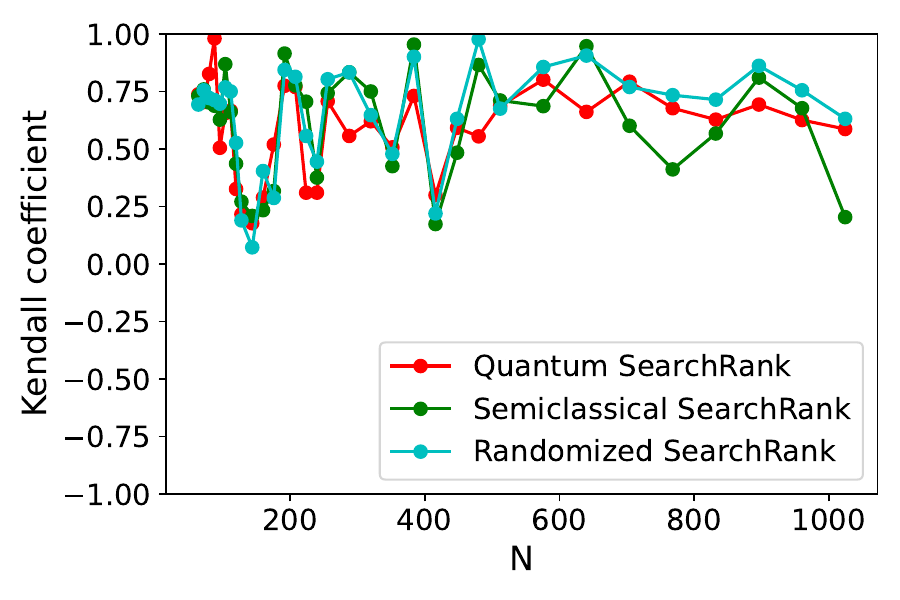}\label{F:kendall_M_48_classical}}
	\subfigure[]{\includegraphics[scale=0.5]{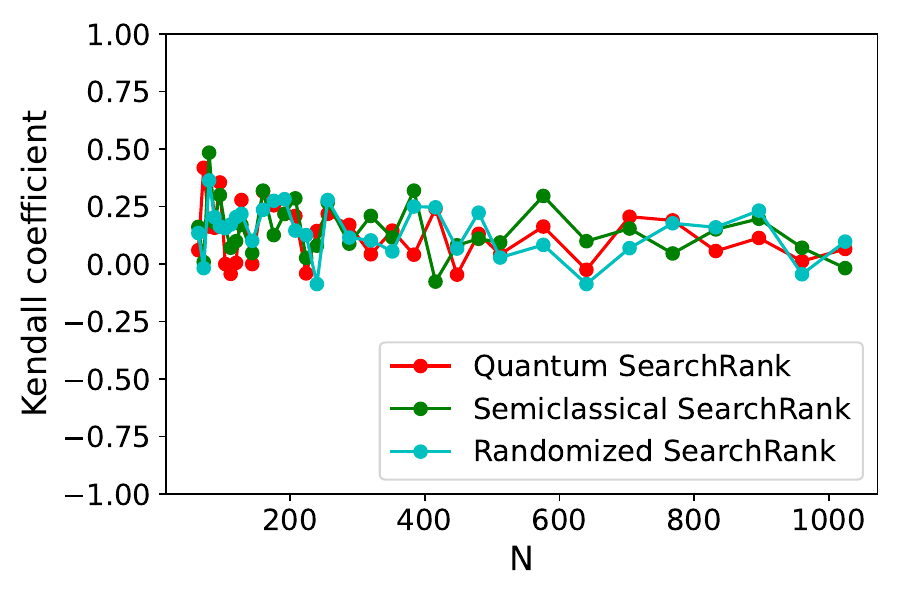}\label{F:kendall_M_48_quantum}}
	\caption{Kendall coefficient versus graph size $N$ for a ranking of $M = 48$ marked nodes by the three SearchRank algorithms compared to a) the classical PageRank and b) the quantum PageRank.}
	\label{F:kendall_M_48}
\end{figure*}

First, let us look at the value of Kendall's coefficient when $M = 48$ nodes are marked. We have plotted this metric for all graphs of different size $N$ in Figure \ref{F:kendall_M_48_classical} for comparison with the classical PageRank and in Figure \ref{F:kendall_M_48_quantum} for the quantum PageRank. The first thing we notice is that the results do not depend on the size of the network $N$. Compared to the classical PageRank, the three SearchRank algorithms have a Kendall coefficient of around $0.6$. Since this coefficient is in the interval $[-1, 1]$, it means that there is a good agreement in the ranking of the marked nodes compared to the ranking of the classical PageRank. However, when compared with the quantum PageRank, this coefficient has a small value, around $0.15$. Therefore, the correlation with the quantum PageRank, although positive, is very weak. As for the best matching SearchRank algorithm, in both cases it seems to be the Randomized SearchRank. Nevertheless, the differences between the three algorithms are practically negligible.

\begin{figure*}[htpb]
	\subfigure[]{\includegraphics[scale=0.5]{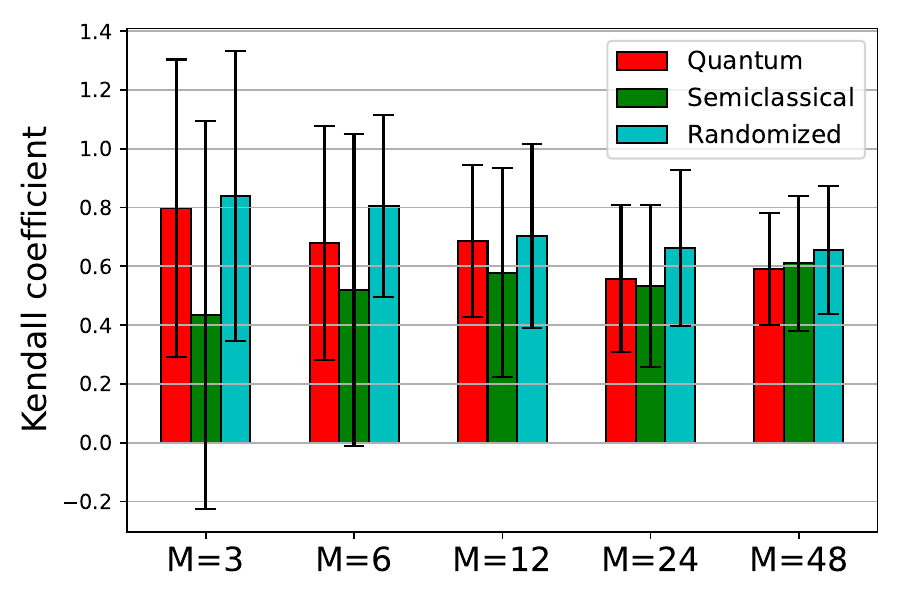}\label{F:kendall_classical}}
	\subfigure[]{\includegraphics[scale=0.5]{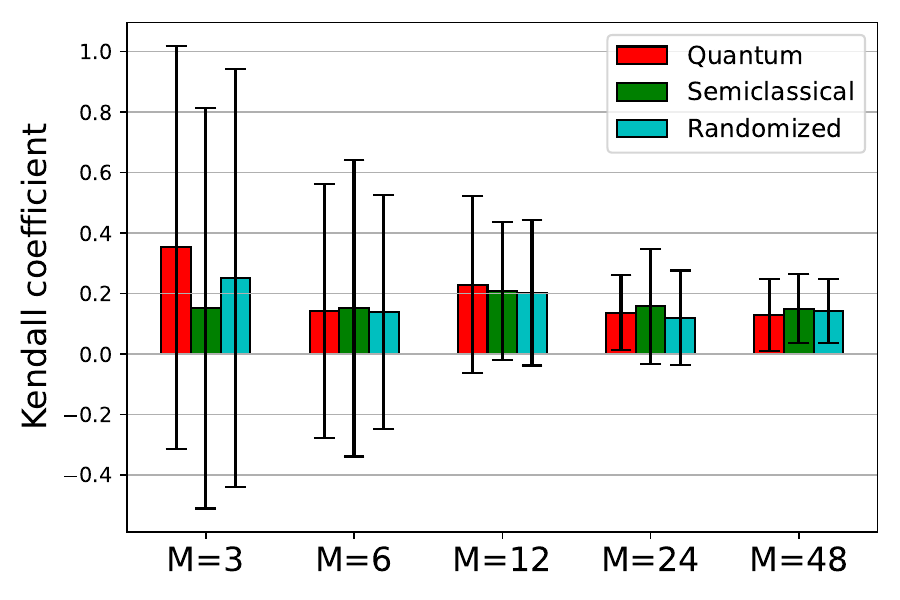}\label{F:kendall_quantum}}
	\caption{Mean Kendall's coefficient for all graphs of different size $N$ in the comparison of the ranking of different number $M$ of marked nodes by the three SearchRank algorithms with a) the classical PageRank and b) the quantum PageRank. Error bars correspond to one standard deviation.}
	\label{F:kendall}
\end{figure*}

Now we want to ensure that these results holds for any number $M$ of marked nodes. We have averaged the Kendall's coefficient for all networks of different size $N$ and plotted it for each value of $M$, in Figure \ref{F:kendall_classical} for the comparison with the classical PageRank and in Figure \ref{F:kendall_quantum} for the quantum PageRank. As expected, in all cases the Kendall's coefficient is larger in the comparison with the classical PageRank, and there is little correlation with the quantum PageRank for all numbers of marked nodes. This may be due to the fluctuations introduced by the quantum PageRank and SearchRank algorithms, so that nodes with similar importance easily change their ranking between the different algorithms. Let us now look at the results for different values of $M$. For $M = 12$, $24$ and $48$ there is little difference between the three SearchRank algorithms in the comparison with the classical PageRank, and the results are similar for all three values of $M$. Nonetheless, for $M = 3$ and $M = 6$ larger differences are observed, with higher values of the coefficient for the Randomized SearchRank. In the comparison with the quantum PageRank all results are almost similar.

It turns out that Kendall's coefficient is quite unstable for small lists. For example, for $M = 3$ it is easy to get perfect agreement, since there are only 3 nodes. However, it is also easy to have perfect disagreement. Thus, due to the fluctuations introduced by quantum evolution in the SearchRank algorithms, we have very different values for each network. This explains the large error bars for small values of $M$.

In summary, although our Semiclassical SearchRank algorithm is not able to sample the quantum PageRank distribution of the marked nodes, it has a good agreement with the classical PageRank. Therefore, this algorithm can be used to sample the marked nodes in a network from a probability distribution that is related to the classical PageRank, taking advantage of quadratic quantum acceleration.

\section{Dependence of the SearchRank Algorithms with the Damping Parameter}\label{Alpha}

\begin{figure*}
	\makebox[0pt][c]{
		\subfigure[]{\includegraphics[scale=0.6]{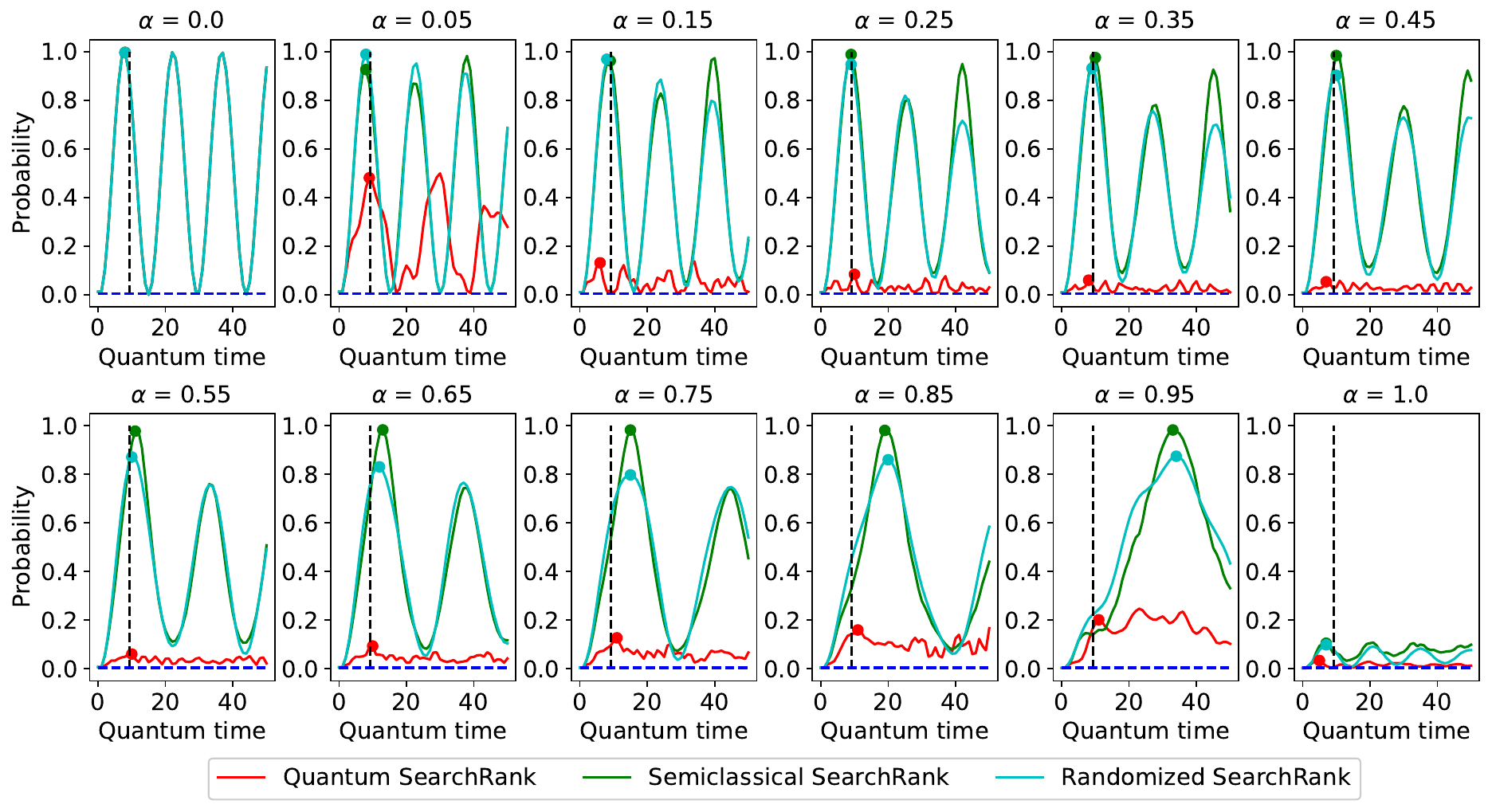}\label{F:alpha_curve}}
	}
	\\
	\vspace{-10pt}
	\makebox[0pt][c]{
		\subfigure[]{\includegraphics[scale=0.572]{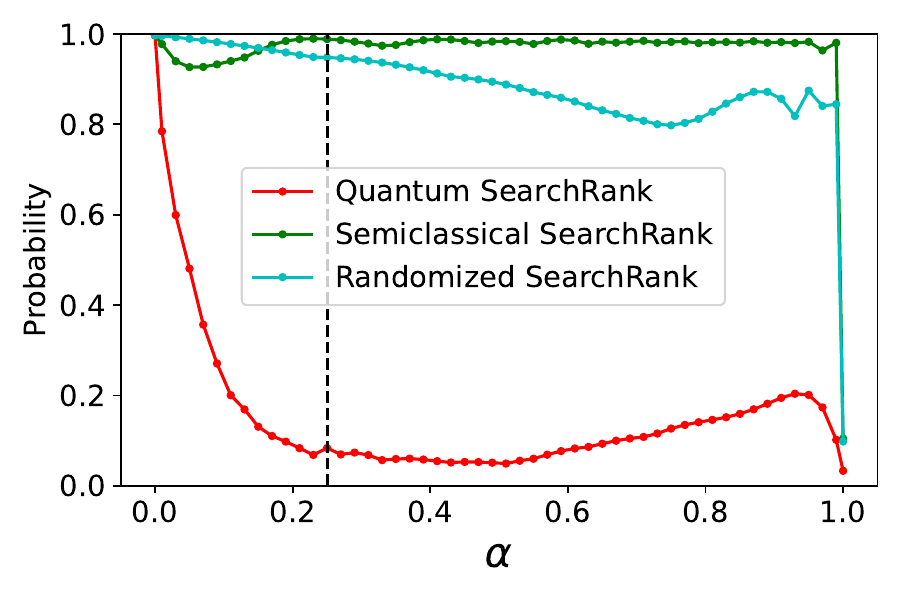}\label{F:alpha_probability}}
		\hspace{+6pt}
		\subfigure[]{\includegraphics[scale=0.56]{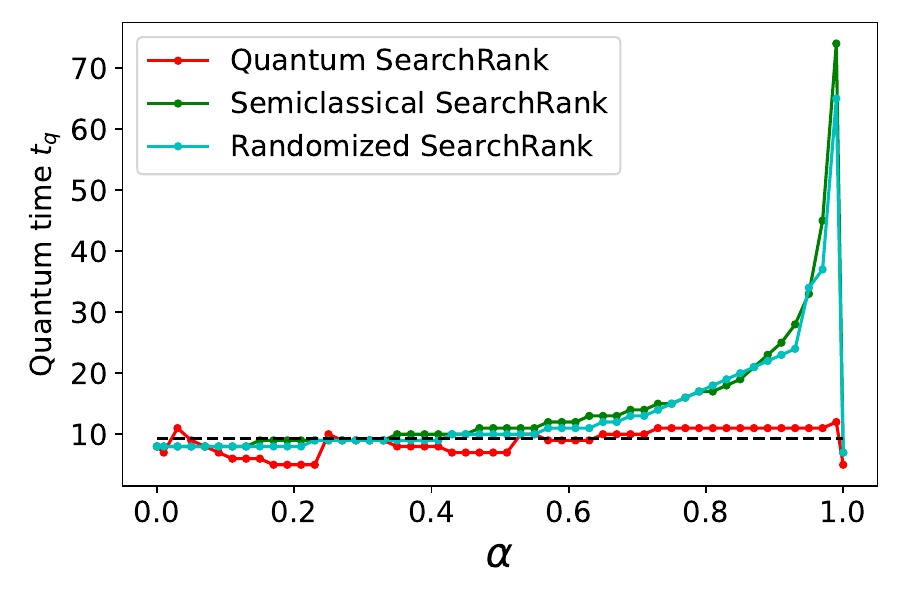}\label{F:alpha_time}}
		\hspace{+2pt}
	}
	\caption{a) Probability curves of measuring one of the marked nodes versus the quantum time for different values of the parameter $\alpha$ for the three SearchRank algorithms applied over a graph with $N=512$ nodes and $M=6$ marked nodes. The first maximum of each curve is marked with a dot. The vertical dashed line represents the reference time $\sqrt{N/M}$. The horizontal dashed lines represent the probability of the marked nodes in the classical (black) and quantum (blue) PageRank distributions. b) Probability achieved at the maximum versus the parameter $\alpha$. The vertical dashed line represents the value $\alpha=0.25$. c) Optimal quantum time for the measurement versus the parameter $\alpha$. The horizontal dashed line represents the reference time $\sqrt{N/M}$.}
	\label{F:alpha}
\end{figure*}

In Section \ref{quantum_searchrank} we stated that the quantum SearchRank used $\alpha = 0.25$. However, the question arises as to what happens if we use another value of $\alpha$. In this section we briefly show some simulation results of the SearchRank algorithms for different values of $\alpha$. We have used a scale-free network with $N = 512$ nodes and $M = 6$ marked nodes.

In Figure \ref{F:alpha_curve} we have plotted the probability of measuring the marked nodes versus the quantum time for different values of the parameter $\alpha$ from $0$ to $1$. Let us first analyze the probability at the first maximum, shown in Figure \ref{F:alpha_probability}. For $\alpha = 0$, when the network is actually the fully connected uniform network given by matrix \textbf{1} in \eqref{G}, the probability at maximum reaches $1$ for all three SearchRank algorithms. Nevertheless, as soon as $\alpha$ takes a non-zero value, the probability in the quantum SearchRank starts to drop dramatically. There is a subtle recovery from $\alpha = 0.5$, but the probability remains below $0.2$ and eventually drops to $0.03$. In the Semiclassical SearchRank the probability at the maximum remains close to $1$ for all $\alpha$ except $\alpha = 1$, when the network is just the scale-free network previous to the mixing made in \eqref{G}. The Randomized SearchRank also maintains a high probability, but drops slowly to $0.8$ in the worst case. Finally, for $\alpha=1$ the probability is almost zero for the quantum SearchRank, and around $0.1$ for the Semiclassical and Randomized SearchRank algorithms.

It might seem that there could be better values than $\alpha = 0.25$ for the semiclassical algorithms, since a large $\alpha$ value means a closer resemblance of the Google $G$ matrix to the original network. However, let us look at the quantum time required to reach the maximum, shown in Figure \ref{F:alpha_time}. For the quantum SearchRank, the maximum is reached at approximately the same time, which is compatible with the reference time $t_q = \sqrt{N/M}$. Nonetheless, for the Semiclassical and Randomized SearchRank, the optimal time increases dramatically from $\alpha = 0.6$. Thus, we cannot increase the value of $\alpha$ much for these algorithms even though the probability remains high. We have found similar results for different examples. Nevertheless, more rigorous analysis is needed in the future.

Interestingly, when $\alpha = 1$, or has a value close to it, the probability of the Semiclassical SearchRank drops so drastically. This suggests that mixing with the fully connected graph to form the Google matrix in \eqref{G} is essential for search. This is therefore a valuable technique not only for the PageRank algorithm, but also for search algorithms with quantum computers. This could lead to new quantum search algorithms inspired by PageRank.

\section{Conclusions}\label{Conclusions}

We have introduced a new quantum algorithm dubbed Randomized SearchRank and compared its performance with other SearchRank algorithms. The quantum SearchRank algorithm is able to expand the probability of measuring a given set of marked nodes, while providing a ranking of these nodes by importance. However, the probability of measuring the marked nodes decreases as the size of the graph increases, so the algorithm loses its utility.

In order to resolve this issue, we have revised Szegedy's semiclassical walk \cite{Semiclassical} formulation and introduced it into the quantum SearchRank. This has resulted in the Semiclassical SearchRank algorithm. Although this algorithm solves the probability problem, it has a longer run time than the quantum SearchRank due to a combination of quantum and classical steps until it converges. Thus, we have proposed a simplification with a single classical step, called Randomized SearchRank, since the underlying walk is equivalent to the quantum walk but with a mixed initial state. Therefore, this new algorithm maintains a similar run time to the original quantum SearchRank.

We have analyzed the performance of the three SearchRank algorithms on a relatively small scale-free network, since such graphs are good models of the World Wide Web. We found that the SearchRank algorithms are able to amplify the probability of the marked nodes, so that there is a high probability of measuring one of these nodes each time the algorithm is run. Since the graph is small enough, the quantum SearchRank does not suffer from probability depletion. Furthermore, the probability of each of the marked nodes produces a ranking with good agreement with the classical and quantum PageRank distributions. However, there is a violation in the ranking of the residual nodes due to the fluctuations introduced by the quantumness of the algorithm, which drastically affect the ranking of nodes with a very small difference in importance.

From this example network, we have been able to visualize how the Semiclassical SearchRank works. In this case, the underlying semiclassical walk behaves like a classical walk on a graph in which the information flow is redirected to the marked nodes, so that the probability of these nodes is amplified in the asymptotic distribution.

To obtain statistical results on the SearchRank algorithms, we have simulated them on a large set of graphs of increasing sizes and different sets of randomly marked nodes. On the one hand, we have checked how the probability of measuring marked nodes in the quantum SearchRank collapses as the size of the graph $N$ increases and/or the number of marked nodes $M$ decreases, with an asymptotic scaling of approximately $\mathcal{O}(M/N)$. This depletion means that the quantum SearchRank loses the ability to amplify the amplitude of the marked nodes, so it is not a successful search algorithm. Nevertheless, we have also shown that in the Semiclassical SearchRank and the Randomized SearchRank this problem is solved, so that the probability does not decrease, remaining at a high value above $0.9$. On the other hand, we have studied the time complexity of the SearchRank algorithms in terms of the quantum time $t_q$. In all cases we have obtained a scaling law approximately compatible with $\mathcal{O}(\sqrt{N/M})$. Since we cannot know a priori what the exact value of the optimal quantum time is, we have decided to take $t_q = \left\lfloor\sqrt{N/M}\right\rceil$ as as a reference value for the measurement, having shown that the probability in the Semiclassical and Randomized SearchRank remains at a high value around $0.9$ despite not being optimal.

Regarding the ranking capability of the SearchRank algorithms, we used Kendall's coefficient to measure the similarity between the rankings provided by the PageRank and SearchRank distributions. We have observed that the three SearchRank algorithms yield similar results, obtaining a fairly good agreement with the classical PageRank. Nonetheless, the agreement with the quantum PageRank is very low, so there seems to be a large lack of correlation. This may be due to the fluctuations introduced by the quantum PageRank, and also by the SearchRank algorithms, so that nodes that are similar in importance can easily have their order changed, and thus the correlation is lost.

Finally, we have studied the dependence of the SearchRank algorithms on the damping parameter $\alpha$. In the case of the quantum SearchRank, the probability of measuring the marked nodes collapses rapidly as soon as $\alpha$ takes a non-zero value. This explains why in the quantum SearchRank a value of $\alpha = 0.25$ was taken \cite{Searchrank} instead of the value of $\alpha = 0.85$ used in the PageRank algorithms \cite{Paparo1}. In the case of the Semiclassical SearchRank, the probability remains close to $1$ except for $\alpha$ values close to $1$. The same is true for the Randomized SearchRank, although the probability drops a little while maintaining a value above $0.8$. Although the probability is high in these last two algorithms, the maximum probability shifts to the right as $\alpha$ increases, so the execution time becomes much higher. Therefore, we cannot increase the value of $\alpha$ arbitrarily. Therefore, $\alpha = 0.25$ seems to be a good value for the new SearchRank algorithms.

Taking all the results together, the Randomized SearchRank solves the probability problem of the quantum SearchRank while maintaining the same time complexity. This algorithm is able to provide one of the marked nodes with a quadratic speedup and a probability that is directly related to the classical PageRank. Thus, it can be used to sample this distribution without the need to calculate it exactly. Moreover, like the quantum SearchRank, it is not necessary to average the results at different time steps of the walk, so it is much faster to implement than the quantum PageRank. It therefore constitutes a further step towards a quantum search engine. Even though the novel Randomized SearchRank stems from the Semiclassical SearchRank, formally the only difference with the quantum SearchRank is that the initial state is a mixed state rather than a quantum superposition of all the $\left|\psi_i\right>$ states. It is interesting to see how the introduction of this mixedness at the beginning of the algorithm allows the probability to be amplified appropriately. Furthermore, blending with the complete graph seems to be crucial for the search functionality. These two intriguing features need further research, and could be used as a base tool for future quantum search algorithms on arbitrary graphs.

In the future, it would be interesting to study other formulations of the unitary Szegedy quantum walk operator with oracles \cite{S_queries} in the context of quantum SearchRank, or extensions with arbitrary phase rotations \cite{APR}. In addition, there are some issues that deserve further research. One of them is the fact that we are assuming that we know the number $M$ of marked nodes to search, so that we can estimate the optimal time for the measurement. However, in a real scenario we would not know how many nodes satisfy the search conditions. A possible solution could be to introduce a quantum counting algorithm \cite{Q_counting_1,Q_counting_2} to estimate the number of marked nodes. Another issue is that we would like to be able to sample from the quantum PageRank distribution as well, since that algorithm is expected to provide better results in quantum networks. In all, our Randomized SearchRank algorithm is already a valuable technique to quantum speed up fundamental properties of classical networks.

\section{Data Availability Statement}

A tutorial for using the python library SQUWALS \cite{Squwals} to simulate the SearchRank algorithms is available on GitHub: \url{https://github.com/OrtegaSA/SQUWALS-repo}.

\section*{Acknowledgments}

We acknowledge the support from the Spanish MINECO grants MINECO/FEDER Projects,  PID2021-122547NB-I00 FIS2021, the “MADQuantum-CM" project funded by Comunidad de Madrid and by the Recovery, Transformation, and Resilience Plan – Funded by the European Union - NextGenerationEU and Ministry of Economic Affairs Quantum ENIA project. M. A. M.-D. has been partially supported by the U.S.Army Research Office through Grant No. W911NF-14-1-0103. S.A.O. acknowledges support from Universidad Complutense de Madrid - Banco Santander through Grant No. CT58/21-CT59/21.

\bibliography{MiBiblio}

\begin{thebibliography}{10}
\bibitem{QT}
{By quantum time we refer to a parameter that determines the duration of the
  quantum evolution, rather than a quantum operator.}
\end{thebibliography}
\bibliographystyle{unsrt}

\end{document}